\documentclass[  ]{aastex631}
\usepackage{graphicx}
\usepackage[outdir=./]{epstopdf}
\usepackage{natbib}
\usepackage{multirow}
\usepackage{xcolor}
\usepackage{newtxtext,newtxmath}
\usepackage{booktabs}
\usepackage{graphicx,subfigure}
\usepackage{multirow}
\usepackage{tabulary,booktabs}
\usepackage{amsmath}
\usepackage{array}
\shorttitle{Galaxy classification}
\shortauthors{Ghaderi, Safari, Alipour}

\begin{document}
\title{Galaxy Morphological Classification with Zernike Moments and Machine Learning Approaches.}

\correspondingauthor{Hossein Safari }
\email{safari@znu.ac.ir}

\author[0009-0005-7934-2752]{Hamed Ghaderi}
\affiliation{Department of Physics, Faculty of Science, University of Zanjan, University Blvd., Zanjan, Postal Code: 45371-38791, Zanjan, Iran.}
\author[0000-0003-3643-5121]{Nasibe Alipour}
\affiliation{Department of Physics, University of Guilan, Rasht, 41335-1914, Iran}
\author[0000-0003-2326-3201]{Hossein Safari}
\affiliation{Department of Physics, Faculty of Science, University of Zanjan, University Blvd., Zanjan, Postal Code: 45371-38791, Zanjan, Iran.}
\affiliation{Observatory, Faculty of Science, University of Zanjan, University Blvd., Zanjan, Postal Code: 45371-38791, Zanjan, Iran.}

\begin{abstract} 
Classifying galaxies is an essential step for studying their structures and dynamics. Using GalaxyZoo2 (GZ2) fractions thresholds, we collect 545 and 11,735 samples in non-galaxy and galaxy classes, respectively. We compute the Zernike moments (ZMs) for GZ2 images, extracting unique and independent characteristics of galaxies. The uniqueness due to the orthogonality and completeness of Zernike polynomials, reconstruction of the original images with minimum errors, invariances (rotation, translation, and scaling), different block structures, and discriminant decision boundaries of ZMs' probability density functions for different order numbers indicate the capability of ZMs in describing galaxy features. We classify the GZ2 samples, firstly into the galaxies and non-galaxies and secondly, galaxies into spiral, elliptical, and odd objects (e.g., ring, lens, disturbed, irregular, merger, and dust lane). The two models include the support vector machine (SVM) and one-dimensional convolutional neural network (1D-CNN), which use ZMs, compared with the other three classification models of 2D-CNN, ResNet50, and VGG16 that apply the features from original images. We find the true skill statistic (TSS) greater than 0.86 for the SVM and 1D-CNN with ZMs for the oversampled galaxy-non-galaxy classifier. The SVM with ZMs model has a high-performance classification for galaxy and non-galaxy datasets. We show that the SVM with ZMs, 1D-CNN with ZMs, and VGG16 with vision transformer are high-performance (accuracy larger than 0.90 and TSS greater than 0.86) models for classifying the galaxies into spiral, elliptical, and odd objects. We conclude that these machine-learning algorithms are helpful tools for classifying galaxy images. \footnote{The Python notebooks are available at \url{https://github.com/hmddev1/machine_learning_for_morphological_galaxy_classification}}

\end{abstract}


\section{Introduction}
 Galaxy morphology encodes key information about the evolutionary path of galaxies, their interactions, and the ongoing physical processes \citep{Baldry2004aip, Noeske2007ApJ, Cano2019MNRAS,li2022automatic,eassa2022automated}.  Combined with physical properties such as colour, redshift, mass,
environmental factors, kinematics, age, and metallicity, it provides a comprehensive
understanding of galaxy populations and their evolution.
The color-morphology relation helps distinguish between star-forming spirals and quiescent ellipticals, while redshift reveals morphological changes over cosmic time. Mass and environment influence the transformation of galaxies, where dense environments lead to a higher fraction of ellipticals. Kinematics and dark matter content relate to galaxies' stability and dynamics. The stellar populations' age and metallicity provide insights into their formation histories.  Together, these properties offer a multidimensional view of galaxies,
helping to unravel the complex processes driving their formation, evolution, and
interaction as a function of cosmic time  \citep{Rezaei2014mnras, Torki2019IAUS..343..512T, Javadi2019IAUS..343..283J}.  In such various descriptions, morphological classification offers prominent insights into the galaxy formation and evolution \citep[e.g.,][]{ Dayal2018PhR, 2024MNRAS.532L..14B}.

Hubble classification is a morphological classification scheme for galaxies \citep{1926ApJ....64..321H}. This scheme classifies galaxies into three main types: spiral, elliptical, and irregular, which use characteristics of arms, discs, bulges, ellipticity, and magnitude. Spiral galaxies include disc-like arms that are mostly younger and have a more active star formation due to a lot of gas and dust in their arm structures. Elliptical galaxies have featureless elliptical shapes that are typically older, and their smooth shapes suggest that they have settled in a more relaxed state with no significant star formation due to the environment's lack of gas and dust. However, irregular galaxies often show evidence of recent measures or interactions with chaotic forms, showing that their structure is still in the reforming stage. de Vaucouleurs extended Hubble's classification for more details in the spiral structure and bar prominence \citep{1976RC2...C......0D}. The Yerkes classification \citep{1957PASP...69..291M}, and David Dunlap Observatory \citep{1970Natur.225..503V} are the later classification systems.

In recent decades, volunteer-driven (citizen science) projects such as GalaxyZoo have encouraged the public to participate in classifying galaxies by analyzing their appearance characteristics \citep{2008MNRAS.389.1179L,2011MNRAS.410..166L, Simmons2017MNRAS,  Lingard2020ApJ}. As technology advanced, parametric and non-parametric fitting techniques were introduced to classify galaxies. Parametric fitting involves modeling a galaxy's light distribution with mathematical functions like the S\'{e}rsic profile to describe structural properties such as size, magnitude, half-light radius, ellipticity, S\'{e}rsic index, and inclination \citep{Balcells2003ApJ, Simard2011ApJS}. The non-parametric approach relies on statistical measures of a galaxy's light distribution, such as CAS (\citep[e.g.,][]{Bershady2000AJ....119.2645B, Conselice2000ApJ...529..886C, Conselice2003ApJS}; concentration, asymmetry, smoothness), Gini coefficient, M20, and clumpiness to describe its structure \citep[e.g.,][]{2003ApJ...588..218A,Cano-D2019MNRAS,2022yCat..74790415A}. Non-parametric techniques are particularly effective for analyzing galaxies with irregular or complex shapes that do not fit well into conventional parametric models.

In recent decades, with widely developing group and space-based instruments, a paradigmatic shift has happened in the astronomy discipline to big data systems \citep{2020A&A...643A.177A}. The various surveys provide big data sets that make it challenging to classify galaxies manually. Then, learning algorithms have been developed to classify galaxy morphology automatically.

The automatic classification of galaxy images has been vastly developed \citep{Shamir2009MNRAS, Fang2023AJ}. Statistical learning has been widely investigated for the morphological classification of galaxy images, increasing the number of astronomical data records and high-process machines that have occurred. \citet{2010MNRAS.406..342B} executed an artificial neural network learning algorithm for morphological classification. They used about 75,000  SDSS DR7  \citep{Abazajian_2009} galaxies. \citet{freed2013application} performed the support vector machine (SVM) algorithm to classify galaxies into spiral, elliptical, and irregular samples.  \citet{Dieleman2015MNRAS} used the translational, rotational symmetry property, and convolutional neural network for the morphological classification of galaxies in the GalaxyZoo project.  They demonstrated that the output of the machine learning algorithm closely matched the consensus of GalaxyZoo participants' majority of questions with high accuracy.  \citet{Selim2017EA} presented a learning algorithm to classify galaxy images via the non-negative matrix factorization,  which is a pattern recognition method and feature selection in image processing \citep{lee1999learning}. They classify the galaxies into elliptical, spiral, and irregular from  the Extraction de Formes Idealisées de Galaxies en Imagerie (EFIGI) catalog. \citet{Cheng2020MNRAS.493.4209C} compared the CNN, k-nearest neighbor, logistic regression, Support Vector Machine, Random Forest, and Neural Networks classification machine learning methods by using $\sim$2800 Dark Energy Survey (DES) data combined with visual classifications from the GalaxyZoo1 project (GZ1). They reached an accuracy of $\sim$0.99 for the morphological classification of elliptical and spiral galaxies. Their study also confirmed a 2.5 percent misclassification of galaxies by GZ1. \citet{Ferreira2020ApJ...895..115F} classified mergers using CNNs and simulated galaxies and measured galaxy merger rates up to redshift $z\sim3$ in CANDELS fields (UDS, EGS, GOODS-S, GOODS-N, COSMOS). They used a Bayesian optimization process over the range of possible hyperparameters to select the deep learning architecture and achieved 90 percent accuracy in classifying mergers from the simulation.  In order to classify galaxies from GalaxyZoo and SDSS,  \citet{REZA2021100492} used
artificial neural networks and the ExtraTrees algorithm. The latter is an ensemble-based
machine learning model that employs extreme randomization to minimize overfitting  \citep{Geurts2006}. 
Their approach achieved an accuracy of over 97.5 percent in classifying sub-images of spirals, ellipticals, mergers, and stars. \citet{li2022automatic} investigated a multiscale convolutional capsule network to extract hidden features of GalaxyZoo2 samples for classifying galaxies.    \citet{Shen2023PASP} proposed a momentum contrastive learning algorithm, which leverages a dynamic memory bank to enhance feature representation and improve classification, achieving 90 percent accuracy for samples collected by GalaxyZoo.   \citet{Fang2023AJ....165...35F} developed rotationally-invariant supervised machine learning, which utilizes adaptive polar-coordinate transformation to standardize galaxy orientations and improve classification consistency, categorizing galaxies into five classes: unclassifiable, irregulars, late-type discs, early-type discs, and spheroids. \citet{Mukundan2024MNRAS} applied the k-nearest neighbors on the non-parametric features (concentration, asymmetry, smoothness, Gini, M20, multimode, intensity, deviation, shape asymmetry, and ellipticity asymmetry) to classify galaxy morphology in SDSS. \citet{Wei2024AJ} composed a layer attention and deformable convolution into a CNN algorithm to morphological categorization as  round, in-between, cigar-shaped, edge-on, spiral, irregular, and error  for GalaxyZoo DECaLS. \citet{Baumstark2024A&C} used simple non-parametric statistics concentration, asymmetry, and clumpiness from SDSS images to classify spiral and elliptical galaxies by random forest algorithm. Numerous machine-learning algorithms have been employed to classify galaxies based on parametric, non-parametric, and other extracted features \citep{10.1111/j.1365-2966.2004.07429.x,Huertas2008A&A, 10.1111/j.1365-2966.2009.15366.x, Malek2013A&A, 10.1093/mnras/stx2351,10.1093/mnras/stx2976,10.1093/mnras/sty503}.

  The parametric properties of galaxies (e.g., magnitude, half-light radius,  S\'{e}rsic index axis ratio, and position angle), as well as the non-parametric (e.g., concentration, asymmetry, smoothness, clumpiness, Gini, and M20) and additional subfeatures (e.g., bulges, discs, spiral arms, bars, rings, and lenses) are essential elements not only to classify the galaxy population into the general classes (early and late type, irregulars), but also to catch its diversity in substructural properties.
 \citet{Tarsitano2018MNRAS}  created  a Dark Energy Survey (DES) Y1 catalog based on galaxy morphology and structure for more than 45 million objects.    The machine learning algorithms utilized the various features discussed above to classify galaxy images.   Their performance and the computational cost  depend on the variety of features and their selection rules. In particular, feature selection is crucial for galaxy classification based on the machine learning algorithms.  Selecting the most relevant features for learning models may improve the performance. Choosing only the necessary features for models can reduce the complexity. Picking fewer features prevents overfitting, noises, and less computational power and memory \citep{Dalal2005IEEE, Hanchuan2005IEEE, Jogin2018inproceedings}. So, a similar question arises about the effect of feature selection (total of parametric and non-parametric measurements and samples) for performance, complexity, overfitting, and efficiency of the machine learning-based classification of galaxy images. Choosing the rotational and translation invariances algorithms is another aspect of galaxy classification due to the variety of perspective angles (orientation), distances, and scales for more than hundreds of billions of galaxies in the observational universe. The data augmentation algorithms (e.g., rotating images during training to enhance rotational invariances) in the convolutional neural network with transformers is a solution to achieve robustness in the machine learning classification of galaxies \citep{Dieleman2015MNRAS, Fang2023AJ....165...35F}. \citet{Tarsitano2022MNRAS.511.3330T} applied elliptical isophote analysis to derive one-dimensional features \citep{Bradley2020zndo...4044744B}, which were subsequently employed to train a boosted tree-based machine learning algorithm for classifying galaxy samples from DES DR2   \citep{Abbott2021ApJS}.
This dimensionality reduction (converting two-dimensional images to a one-dimensional set of features) is an essential solution to achieve less computational cost for classifying astronomical objects. They obtain 86 and 93 percent classification accuracy for early and late types galaxies, respectively. 

 The well-known parametric and non-parametric features are essential quantities that introduce the different types of galaxies. Suppose the sample images of two same type observed galaxies; we expect similar parameters, but these two samples show minor differences in the fine structures. This implies that the parameters are not unique to describe these galaxy images. Therefore, using a set of parametric and non-parametric quantities is not a unique feature used to describe the galaxy images.

Describing a two-dimensional image (original image) based on a set of complete and orthogonal basis functions is essential to finding unique features. Applying the inverse transformation to these features gives the reconstructed image, ensuring the uniqueness of features. Here, we introduce Zernike polynomials (ZPs) as a complete and orthogonal basis to describe a two-dimensional image. The ZPs are orthogonal complete sets and are defined on the unit circle, meaning any image can be expressed as a linear combination of ZPs \citep{Mukundan19951433, Niu2022JOpt...24l3001N, SafariIJJA2023}. The Zernike moments (ZMs) are the unique features calculated for an image to describe its characteristics. ZMs are the coefficients of this linear combination of an image based on the ZPs \citep{Teague1980, Khotanzad1990, alipour2012, Javaherian2014}.  Due to the characteristics of ZPs, the ZMs are unique and independent, and their magnitudes are rotation invariant. Also, the original image can be reconstructed with a minimum error using a finite set of ZMs and the inverse transformation method. These properties are benefits of using the ZMs to describe an image to apply for machine classification algorithms  in many disciplines such as fingerprint recognition \citep{Kaur2019}, detecting eye in facial images \citep{Kim2008ETRI}, diagnosing the brain tumor \citep{ZHENG2023Biomedical}, predicting protein-protein interactions from protein sequences \citep{Wang2017ijms}, and recognizing alphabets in a text \citep{BROUMANDNIA2007717}. Recently, the composition of ZMs and SVM applied for identification and tracking solar brightening and dimming features \citep{rad2012large,alipour2015, Honarbakhsh2016SoPh, Yousefzadeh2016SoPh,hosseini2020, shok2022, Alipour2022, Shokri2024}, the prediction of flaring active regions \citep{Raboonik2017ApJ,alipour2019}, and classification of radio galaxies \citep{Sadeghi_2021}. 

In this work, we   collect  a total of 11,735 galaxies and 545 non-galaxies images from GalaxyZoo2 (GZ2), applying the threshold task (features or disc fraction, edge-on no fraction, spiral fraction, smooth fraction, completely round fraction, odd no fraction, and odd yes fraction). First, we classify the GZ2 images into the galaxy and non-galaxy classes. To do this, we apply the five classification models that used the ZMs (extracted from GZ2 images) in support vector machine (SVM) and convolutional neural network (CNN) as well as the features extracted from original images in the  2D-CNN, ResNet50, and VGG16 with vision transformer algorithms. Second, we  design models similar to the above to recognize galaxy images in spiral, elliptical, and odd objects (e.g., ring, lens, disturbed, irregular, merger, and dust lane) classes. We  measure  the performances of classification algorithms based on accuracy, precision, recall, $f_{1}$-score, and true skill statistic (TSS) metrics. The area under the curve (AUC) in the receiver operating characteristic (ROC) curve is measured by the classifiers' performances compared with the random classification model.

Sections \ref{data} and \ref{methods}  describe the collected  galaxies and non-galaxies  images and the analysis methods, respectively. Section \ref{results} gives the Results. Section \ref{disc} explores the Discussions and comparisons with the previous studies. Section \ref{con} makes a concluding remarks of this article.

\section{Data}\label{data}
GalaxyZoo2  \citep{Willett_2013,2016Hartmnras}  collected  more than 304,000 galaxy images from the   Sloan Digital Sky Survey (SDSS)  \citep{2002StraussAJ} based on a citizen science project (visual inspection) recognizing the morphological structure of samples.  We collect the population of each class from  gz2$\_$hart16 (a CSV table available at \url{https://data.galaxyzoo.org/#section-7}) by applying    threshold cuts for each class.  The  gz2$\_$hart16 table tabulates objectID numbers from SDSS data release 7 (column 1), the locations in declination and right ascension (columns 2 to 5), gz2$\_$class for the most common consensus morphology (column 7), total$\_$votes (column 9), and the eleven tasks (questions) and volunteer responses (columns 10 to 232). For each response of tasks in the gz2$\_$hart16 table, the fraction represents the percentage of volunteers for specific classification answers of a given galaxy. \citet{Willett_2013} represented the complete list of tasks and responses to questions (Figure 1 and Table 2 therein).

Applying the "star or artifact fraction" task in the GZ2 catalog indicated by ‘A’ gives about 545 non-galaxy samples. The letter ‘A’ in the list of gz2$\_$hart16 table indicates the non-galaxy images, including stars and artifacts features. We select 11,735 galaxy samples with thresholds for the fraction task, including spiral, elliptical, and odd object classes. We use "features or disc fraction", "edge-on no fraction", and "spiral fraction" thresholds higher than 0.95 to collect the spiral samples. These tasks give 6,139 clean samples in the spiral galaxies class. The threshold of 0.9 for "smooth fraction", "completely round fraction", and "odd no fraction" tasks return 4,077 clean samples of the elliptical galaxy class. Setting the "odd yes fraction" task higher than 0.9 provides 1,519 samples in the odd objects class, including ring, lens, disturbed, irregular, merger, dust lane, and other galactic features.
 
\section{Methods}\label{methods}
Here, the SVM and CNN are two classifiers that used the original galaxy or non-galaxy images or their ZMs. In the remainder of this section, we briefly  discuss the properties of ZMs, characteristics of the SVM classifier, and CNN as the building blocks of the classifier machine to categorize images into galaxies (spiral, elliptical, and odd objects) and non-galaxies  (stars, artifacts, cosmic rays, image noise, etc.).

\subsection{Zernike moments}\label{sec:ZM}
The Zernike polynomials (ZPs) are   an orthogonal complete set  of two-dimensional complex functions that are defined in a unit circle $(x^{2}+y^{2} \leq1)$. ZPs with order $(p)$ and repetition $(q)$ in the polar coordinate $(r,\theta)$ are defined as:

\begin{equation}
    ZP_{p}^{q}(r,\theta) = R_{p}^{q}(r)\,\exp(iq\,\theta),
    \label{zps}
\end{equation}

where radial function $(R_{p}^{q}(r))$ is given by:

\begin{equation}
     R_{p}^{q}(r) = \sum_{k=0}^{\tfrac{p-q}{2}} \frac{(-1)^k\,(p-k)!}{k!\left (\tfrac{p+q}{2}-k \right )! \left (\tfrac{p-q}{2}-k \right)!} r^{p-2k},
     \label{Vpq}
\end{equation}

in which  $p-q=$ even and $|q|\le p$.  ZPs are orthogonal in a unit circle of polar coordinate as, 

\begin{eqnarray}
 \int_{0}^{2\pi}{\int_{0}^{1} ZP_{p}^{* q} ZP_{p'}^{q'} r dr d\theta} = \frac{\pi}{p+1} \delta_{pp'}\delta_{qq'},
  \end{eqnarray}
  
 where $\delta$ indicates the Kronecker delta function. The $ZP_{p}^{* q}$ is the complex conjugate of $ZP_{p}^{q}$.

\begin{figure}
\centering
\includegraphics[width=0.7\textwidth] {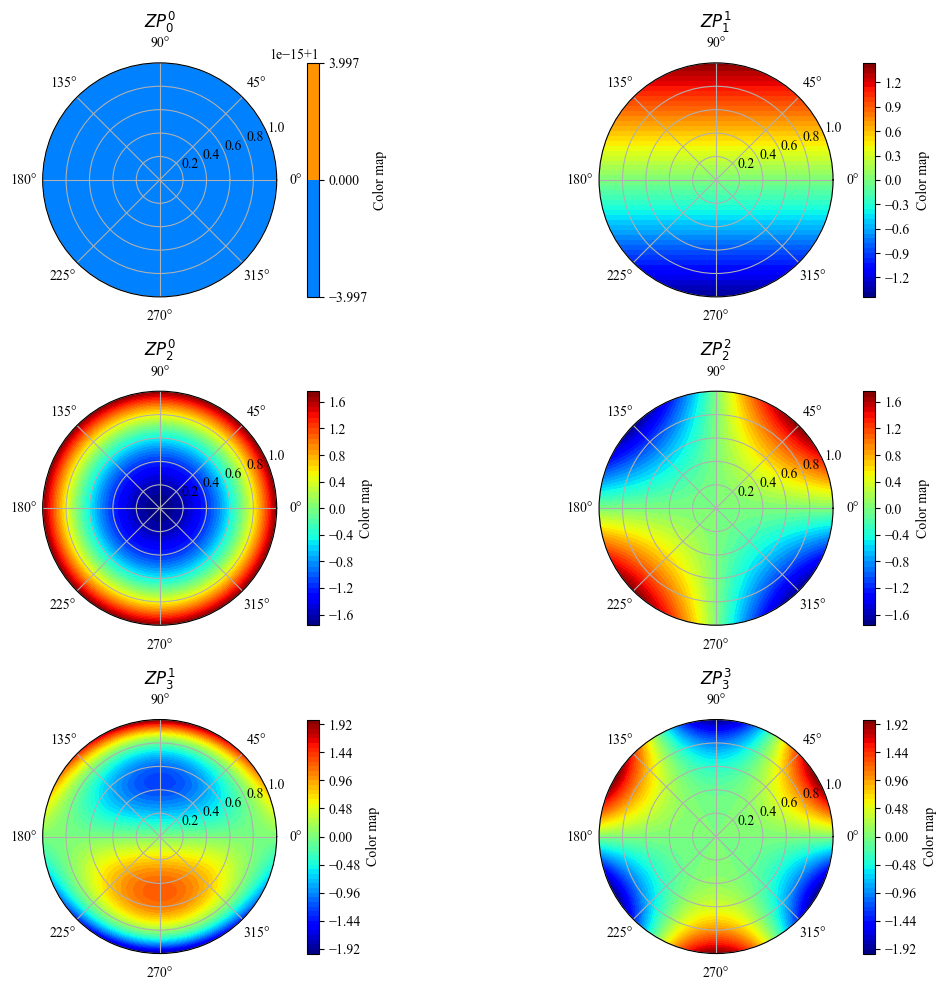}
       \caption{  The Zernike polynomials (ZPs) for different order numbers $p = 0, 1, 2,$ and $3$ in polar coordinate.}
       \label{zpolar}
\end{figure}

Figure \ref{zpolar} shows the ZPs for different order numbers of $p = 0, 1, 2,$ and $3$ in the polar coordinate.
The ZPs with $p=0$ ($ZP_{0}^{0}$=1) indicate the constant of the set, $p=1$ ($ZP_{1}^{-1}$ and $ZP_{1}^{1}$) corresponds to tilt in the horizontal and vertical directions. The ZPs for $p=2$ ($ZP_{2}^{-2}$, $ZP_{2}^{0}$, and $ZP_{2}^{2}$) show defocus (focus shift) and astigmatism (different  focal lengths  in two orthogonal directions). Also, ZPs with $p=3$ ($ZP_{3}^{-3}$, $ZP_{3}^{-1}$, $ZP_{3}^{1}$, and $ZP_{3}^{3}$) correspond to coma asymmetrically aberration and trefoil. The higher order ZPs describe the complex aberrations to characterize wavefront errors in optical systems, helping to improve distortions and optimize imaging quality \citep[e.g.,][]{Zernike1934MNRAS..94..377Z, Noll1976JOSA...66..207N, Lakshminarayanan2011JMOp...58..545L, McAlinden2011Clinical}.

The ZPs in lower order numbers ($p \leqslant 2$) capture broad global features in an image. The order numbers ($3 \leqslant p \leqslant 5$) identify finer details, such as asymmetries or specific distortions in an image. While the higher order numbers ($p \geqslant 6$) extract the tiny features like lines, sharp edges, and high-frequency variations due to the oscillatory behavior of the radial component of higher ZPs. Therefore, we can describe the details and structures of objects in an image by applying a series of ZPs in the polar coordinates.

The ZPs help describe an image in global and finer details features. Zernike moments (ZMs) are coefficients derived from ZPs that describe the shape and structural features of an image. ZMs map an image to a series of complex numbers based on ZPs. The Zernike moment for order number $p$ and repetition number $q$ is given by

\begin{eqnarray}
 Z_{p}^{q}=\frac{p+1}{\pi}\int _{0} ^{2\pi} {\int _{0} ^{1} {I(r,\theta) ZP^{* q}_{p} rdrd\theta}},
 \label{Zpq}
\end{eqnarray}

in which $I(r, \theta)$ is the registered intensity of image pixels in the polar coordinate.
The discrete ZMs for a digital image with the size of $M \times N$ is defined by

\begin{eqnarray}
Z_{p}^{q}=\frac{p+1}{\pi} \sum _{i=0} ^{M-1} {\sum _{j=0} ^{N-1} {I(i,j) ZP_{p}^{q}(r_{ij},\theta_{ij})},}
\end{eqnarray}

where $r_{ij}=\sqrt{x_i^2+y_j^2}$ and $\theta_{ij}=\arctan(\frac{y_j}{x_i})$
are the polar coordinate elements of the image in a unit disc.
Since the ZPs are complete and orthogonal set in two-dimensional function space, ZMs are unique and independent features \citep{Khotanzad1990,alipour2019, Alipour2022}. To show the uniqueness of ZMs, one may reconstruct the original image by applying the inverse transformation. The reconstructed image is obtained by: 

\begin{eqnarray}
   I_{R}(r,\theta)=\sum _{p=0}^{p_{\rm max}} \sum _{q} {Z_{p}^{q} ZP_{p}^{q}(r,\theta)}.
   \label{rec}
\end{eqnarray}

 Due to the Fourier (exponential) term in the Zernike polynomials (Equation \ref{zps}), the magnitude of ZMs remains invariant to the rotation \citep{Khotanzad1990}. Suppose $I'(r, \theta')$ is the rotated image intensities with angle $\alpha$ in the anticlockwise direction of the original image $I(r, \theta)$, then we have $\theta' = \theta - \alpha$. Using equation (\ref{Zpq}), we obtain the relation between the ZMs of the rotated and original images as:

 \begin{eqnarray}
    Z_{p}^{' q}=Z_{p}^{q}\exp(-iq\alpha).
    \label{z_rot}
 \end{eqnarray}

\begin{figure}
\centering
\includegraphics[width=0.8\textwidth] {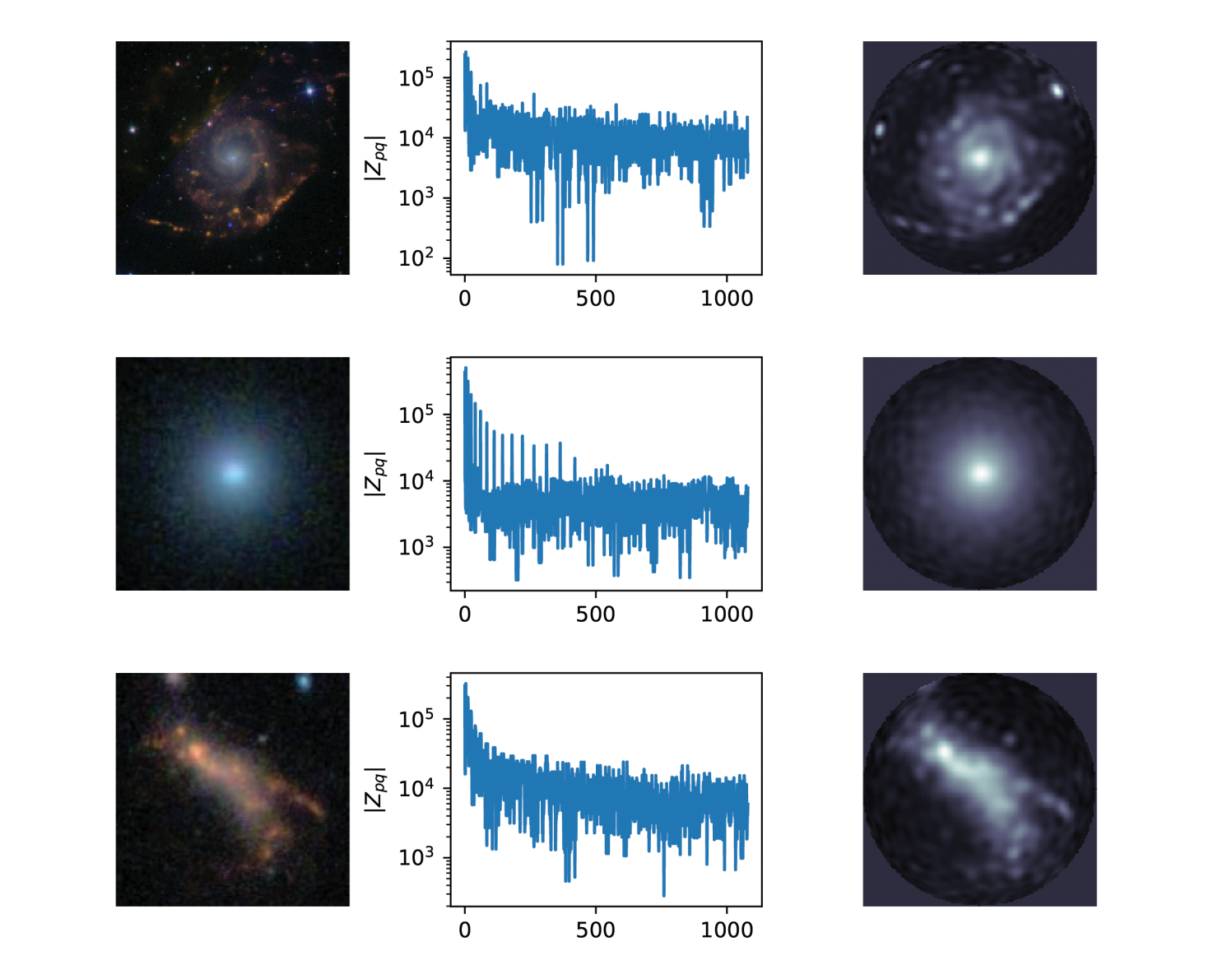}
       \caption{The original image of galaxies recorded by SDSS (left column), the magnitude of Zernike moments (ZMs) for $p_{\rm max}=45$ (middle column), and reconstructed (Equation \ref{rec}) image (right column). The first, second, and third rows correspond to spiral, elliptical, and odd objects, respectively.}
       \label{recon}
\end{figure}

 Equation \ref{z_rot} shows that the ZMs of the rotated and original images differ in a phase factor. Therefore, the magnitude of ZMs for rotated and original images are precisely the same ($|Z_{ p}^{' q}|=|Z_{p}^{q}|$), which employs the rotation invariancy magnitude of ZMs.
 Figure \ref{recon} shows the original images in RGB (left panels), the absolute value of ZMs with $p_{max}=45$ (middle panels), and the reconstructed images (right panels) for a spiral (top row), elliptical (middle row), and odd objects (bottom row) galaxies, respectively. To compute the ZMs for galaxy images in RGB (red, green, and blue) channels, we first convert the RGB image to grayscale using the transformation of $I=0.299R+0.587G+0.114B$ in which I, R, G, and B are the weighted intensity of grayscale, intensity in R (red channel), G (green channel), B (blue channel) for each pixel, respectively. The length of ZMs is 1081   (middle column)  with maximum order number $p_{max}=45$.  The reconstructed images (right column) are obtained using Equation (\ref{rec}). The ZEMO \citep{SafariIJJA2023} is a package to compute the ZMs, and the reconstructed image is available on GitHub (\url{https://github.com/hmddev1/ZEMO}) and PyPI (\url{https://pypi.org/project/ZEMO/1.0.0/}).

\subsection{Support Vector Machine}\label{sec:HVG}
The support vector machine (SVM) is a supervised machine learning algorithm used primarily for two-class (binary) problems \citep[e.g.,][]{jakkula2006tutorial}. The SVM algorithm uses the optimization Lagrangian multiplier to determine the coefficients of the hyperplane in the feature space \citep{cortes1995support}. The optimal hyperplane (decision boundary)  in the high-dimensional feature space separates data points of different classes. Optimizing the optimal hyperplane that maximizes the distance (margin) of different classes (support vectors) is an essential step in the SVM machine learning algorithm. SVM also uses the kernel functions to handle the non-linear data, transforming the data into the high dimensional space in which the linear separation is possible \citep{Scholkopf1998, Hearst1998}. The polynomial and radial basis function (RBF) are standard kernels for SVM. Using the RBF kernel, the SVM effectively classifies the high-dimensional feature spaces and is robust against overfitting if the number of dimensions exceeds the number of samples. The RBF kernel $K(x,y)$ commonly transforms the distance between points $x$ and $y$ in feature space via the Gaussian function as

\begin{eqnarray}
 K(x_i, x_j) = \exp \left( -\gamma \|x_i - x_j\|^2 \right)
 \label{kernel}
\end{eqnarray}

where $\gamma$ defines the width (scale) of the kernel and $\|x_i - x_j\|^2$ represents squared Euclidean distance.

In the context of morphological galaxy classification, SVM can learn a decision boundary that separates distinct types of galaxies by mapping input galaxy data into a high-dimensional feature space using kernel functions. We used the radial basis function (RBF) as the kernel and the penalty parameter C=1.5.   The penalty parameter controls the trade-off between maximizing the margin and minimizing the classification error in the training process \citep{hsu2003practical}. 

\subsection{Deep Learning Algorithm  (classic CNN, ResNet50, and VGG16 with vision transformer)}\label{sec:HVG}

\begin{figure}
\centering
\includegraphics[width=0.7\textwidth] {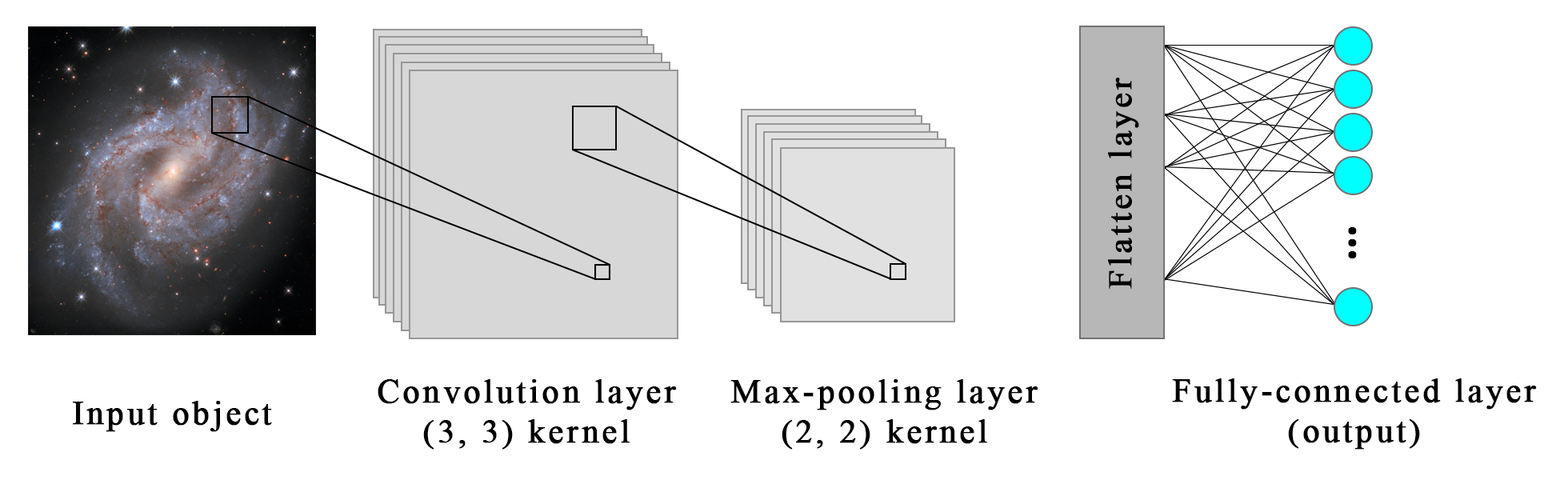}
       \caption{The visualization scheme for the convolutional neural network (CNN) layers includes a convolutional layer with a $3 \times 3$ kernel, a max-pooling layer with a $2 \times 2$ kernel, a flattening layer, and a fully connected layer, applied to a sample galaxy image.
}
       \label{cnnscheme}
\end{figure}

The Convolutional Neural Network (CNN) is a well-developed neural network algorithm in machine learning \citep[e.g.,][]{Simonyan2014arXiv, Shea2015arxiv, LiIEEE2022}. 
 The 1D-CNN, 2D-CNN, ResNet50, and VGG16 use the central convolution of CNN with different architect structures. Figure \ref{cnnscheme} shows the scheme of the CNN layers. The CNNs were developed to use convolutional layers containing filters (kernels) for one-dimensional features or images. These filters contain the learning parameters applied for feature extraction by matrix multiplying algebra (activation mapping) from 1D feature arrays or 2D images. For an image, edges, textures, and patterns are extracted in the  convolutional layers in the CNN algorithm. These training parameters are stored in the depth dimensions. This is the core building block of a CNN. The pooling layers apply the average or max pooling for sub-regions in the activation maps to reduce the number of extracted parameters in the depth layers. It retains the most essential information and reduces the complexity of the overfitting. The flattened layer transforms the multi-dimension features into one dimension, feeding them to a fully connected layer (Dense layer).
The dense layer processes the flattened outputs from previous layers, learning complex relationships between high-level features. In classification tasks, the final dense layer typically outputs a vector of class probabilities, making it essential for decision-making. The dense layer performs the classification step via the learning process following several max-pooling and convolutional layers.

\begin{table}
\caption{The primary characteristics, advantages, disadvantages, and computational costs for SVM, Classic CNN, ResNet50, and VGG16.}
\hspace{2cm}
\centering
\begin{tabular}{|>{\centering\arraybackslash}p{1.8cm}|>{\raggedright\arraybackslash}p{3.5cm}|>{\raggedright\arraybackslash}p{3cm}|>{\raggedright\arraybackslash}p{3cm}|>{\raggedright\arraybackslash}p{3cm}|}
\hline
\textbf{Model} & \textbf{Characteristics      (Layers)} & \textbf{Advantages} & \textbf{Disadvantages} & \textbf{Computational Cost} \\ 
\hline

\textbf{SVM} & 
- No layers, traditional machine learning algorithm \newline
- Uses a kernel (e.g., RBF) to classify data \newline
- Manual feature extraction required & 
- Works well for small datasets \newline
- Simple to implement \newline
- Effective with limited features & 
- Requires manual feature extraction \newline
- Does not scale well for large datasets \newline
- Struggles with high-dimensional data & 
Low for small datasets \newline
High for large datasets \\ 
\hline

\textbf{Classic CNN} & 
- Input layer (Image) \newline
- Convolutional layers \newline
- Max-pooling layers \newline
- Flattening layer \newline
- Dense (fully connected) layers & 
- Automatic feature extraction \newline
- Hierarchical feature learning \newline
- Scalable for larger image datasets & 
- Limited depth (compared to ResNet and VGG) \newline
- May not capture subtle, complex features & 
Moderate \newline
Depends on the number of layers and filters \\ 
\hline

\textbf{ResNet50} & 
- +50 layers \newline
- Convolutional layers \newline
- Residual blocks (skip connections) \newline
- Max-pooling layers \newline
- Dense (fully connected) layers & 
- Deep architecture allows learning complex, hierarchical features \newline
- Solves vanishing gradient problem \newline
- Excellent for high-accuracy classification & 
- High computational and memory requirements \newline
- Slower training time & 
High \newline
Requires powerful GPUs and large memory \\ 
\hline

\textbf{VGG16} & 
- 13 layers \newline
- Convolutional layers with small filters (3x3) \newline
- Max-pooling layers \newline
- Flattening layers \newline
- Dense layers & 
- Pre-trained on ImageNet, useful for transfer learning \newline
- Simple architecture, easy to understand \newline
- Good at feature extraction from images & 
- Large model size (parameters) \newline
- High memory usage \newline
- Lacks residual learning (found in ResNet) & 
High \newline
Requires significant computational resources \\ 
\hline

\end{tabular}
\label{comodels}
\end{table}

The 2D-CNN, ResNet50, and VGG16  apply the Vision Transformer (ViT) in the statistical learning of classification models. The ViT is a deep learning architecture that applies the principles of transformers. It was initially used in natural language processing in computer vision \citep{Dosovitskiy2020arXiv}. The ViT is also an image augmentation approach extracting elaborate features for the training process of classifiers. The vision transformer in the CNN model uses a center crop, random horizontal and vertical flips, random rotation, random resized crop, and normalization in the feature extraction procedure.

Here, we applied the classic CNN, ResNet50, and VGG16 for classifying GZ2 galaxies using the features extracted from the vision transformer algorithm. The classic CNN, ResNet50, and VGG16 are different architects of the neural network with different convolutional, max-pooling, flattening, and dense layers. The ResNet50 and VGG16 are the pre-trained architectures of transfer learning   methods  in the CNN. ResNet50 applies 1$\times$1 convolution, 3$\times$3 convolution, and another 1$\times$1 convolution. The ResNet50 includes four  convolutional  layer stages, with multiple residual blocks for each layer. However, VGG16 consists of 13   convolutional  layers grouped in five different blocks. The blocks include multiple   convolutional  layers and a max-pooling layer.   Table \ref{comodels} represents the main characteristics, advantages, disadvantages, and computational costs of the SVM, classic CNN, ResNet50, and VGG16. Classic 2D-CNN, ResNet50, and VGG16 with ViT are well-known high-performance image classification models that apply automatic feature extractions from two-dimensional images. To investigate the capability of ZMs for describing galaxy images, comparing the performance metrics of machine-learning algorithms of SVM and 1D-CNN with ZMs and automatic feature extraction methods (such as 2D-CNN, ResNet50, and VGG16 with ViT) is a possible solution.

\section{ Results}\label{results}

  First, we describe the properties of galaxy images of spiral, elliptical, and odd objects using the ZMs that are extracted from images based on ZPs. Second, we apply the SVM, 1D-CNN, 2D-CNN with ViT, ResNet50 with ViT, and VGG16 with ViT for the classification of galaxy images from the GZ2 catalog to compare the performance of the classifiers that directly used one-dimensional ZMs (SVM and 1D-CNN) with the classifiers applied the automatic feature augmentation algorithms (2D-CNN, ResNet50, and VGG16 with ViT).
In the remainder of this section, we developed the classifier machine (1) galaxy-non-galaxy and (2) galaxy classifier. The galaxy-non-galaxy classifier recognizes the GZ2 images into galaxy and non-galaxy  (stars, artifacts, cosmic rays, image noise, etc.). The galaxy classifier identifies the type of galaxies into spiral, elliptical, and odd objects main classes. The Python codes and data sets for classifier models are available  on  GitHub (\url{https://github.com/hmddev1/machine_learning_for_morphological_galaxy_classification}).

\subsection{  ZMs for galaxies and non-galaxies }
  First, we investigate the property of ZMs for elliptical galaxies and their relationships with some structural and morphological quantities. By definition, the perfect elliptical galaxies are more uniformly distributed in brightness with smooth, featureless, and elliptical appearances. This property of elliptical galaxies implies that the brightness profile ($I$) is most likely to be axially symmetric, meaning that $I(r,\theta)=I(r)$ in which $r$ and $\theta$ are radius and angle in the polar coordinate, respectively. 
The elliptical galaxy brightness is well-modeled by the S\'{e}rsic profile \citep{Sersic1963BAAA....6...41S}. The S\'{e}rsic profile is given by, 

\begin{equation}
    I(r) = I_e \, \exp \left (-b_n \left[ \left(\frac{r}{r_e}\right)^{\frac{1}{n}} - 1 \right] \right ).
\end{equation}

 where, $I_e$, $r_e$, and $n$ are the effective intensity, effective radius, and S\'{e}rsic index, respectively, and $b_n \approx 2n-1/3$.  For elliptical galaxies, the S\'{e}rsic index ($n$) ranges from 2 to 6. The index $n=4$ indicates the de Vaucouleurs profile typically applied for elliptical galaxies. 

  Using Equation (\ref{Zpq}) and the axially symmetric behavior of the elliptical galaxies, the ZMs vanish for the non-vanishing repetition numbers $q \neq 0$ ($Z_{p}^{q\neq 0}$=0) due to the Fourier terms ($\exp(iq\theta)$) in the ZPs (Equation \ref{zps}). Consequently, the ZMs vanish for the odd order number ($p=2l+1$), satisfying the selection rule for repetition number ($|p-q|$ should be an even number) in the ZPs. The non-vanishing ZMs for even order numbers ($p=2l$) are, 

 \begin{equation}
     Z_{2l}^{0}=2(2l+1) \int _{0} ^{1} {I(r) ZP_{2l}^{* 0} rdr}.
     \label{Z2l0}
 \end{equation}

  Immediately, setting $l=0$ gives,  

\begin{eqnarray}
 I_{\rm tot}=&  \pi Z_{0}^{0}\\
  =& 2 \pi I_e r_e^2 \frac{n e^{b_n}}{(b_n)^{2n}} \Gamma(2n), 
\end{eqnarray}

  in which $I_{\rm tot}$ is the total brightness of a galaxy. Hence, the apparent magnitude ($m$) of the galaxy is introduced by, 

\begin{equation}
    m=-2.5 \log_{10} \pi Z_{0}^{0}+\rm const.
\end{equation} 

\begin{figure}
\centering
\includegraphics[width=0.6\textwidth] {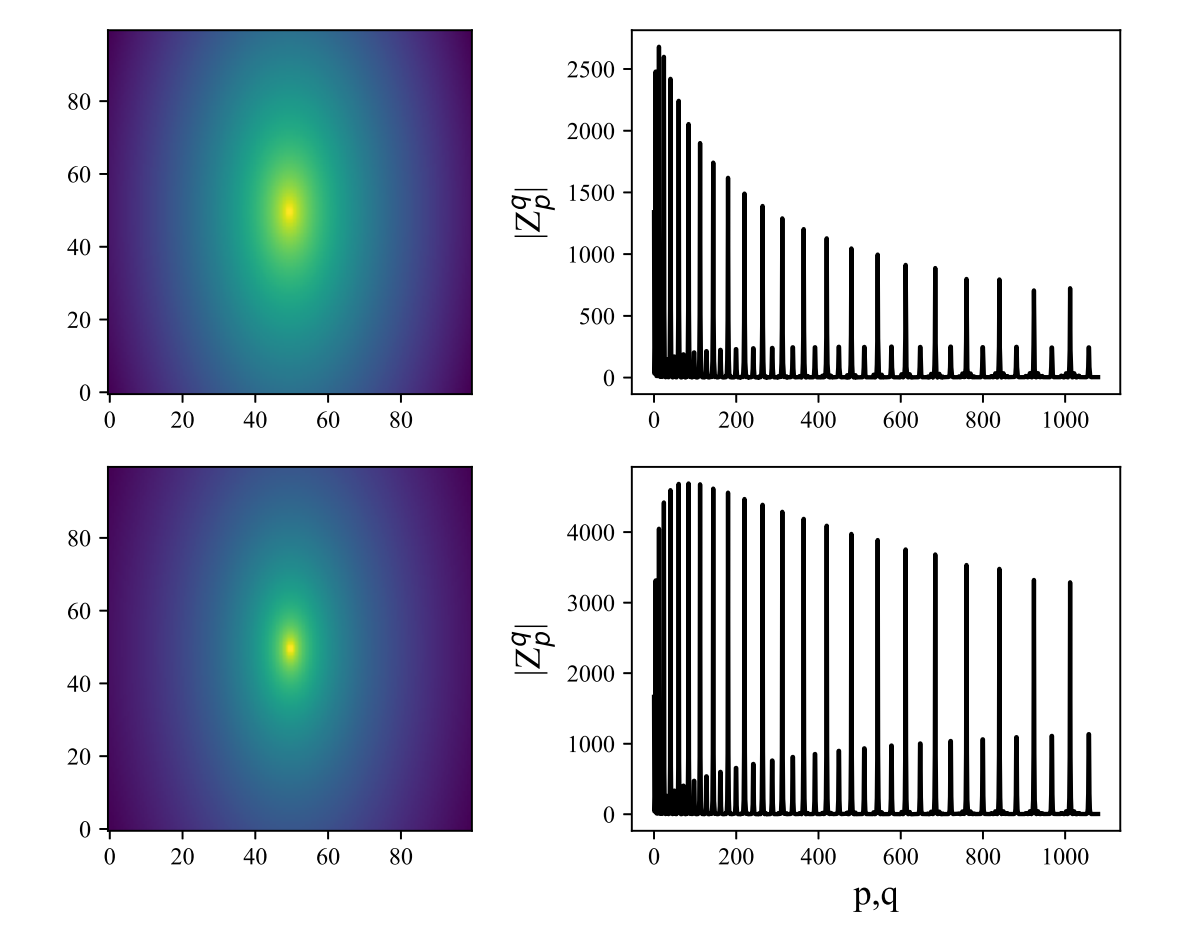}
\caption{ 
To samples of elliptical galaxies (left column) and the corresponding ZMs with $p_{\rm max}=45$ (right column) for S\'{e}rsic index $n=2$ (top row) and $n=4$ (bottom row). The $p_{\rm max}=45$ returns 1081 ZMs with the set of orders and repetitions labeled by $(p,q)$. } 
       \label{ellip_Ser}
\end{figure}

  Figure \ref{ellip_Ser} shows two samples of elliptical galaxy images and the corresponding magnitude of ZMs for S\'{e}rsic index $n=2$ (top row) and $n=4$ (bottom row). We choose the maximum order number $p_{\rm max}=45$, which returns 1081 ZMs. We observe that the elliptical galaxies' brightness shows axial symmetry behavior. Hence, we expect significant peaks to appear for the ZMs with even order numbers ($p = 2l$) and zero repetition ($q = 0$) satisfying Equation \ref{Z2l0} for both $n=2$ and four simulated elliptical galaxies. We also observe some tiny peaks for ZMs with odd order and repetition numbers. These trivial peaks may occur due to the discrete form of the image function instead of being a continuous function in computing ZMs.

\begin{figure}
\centering
\includegraphics[width=0.9\textwidth] {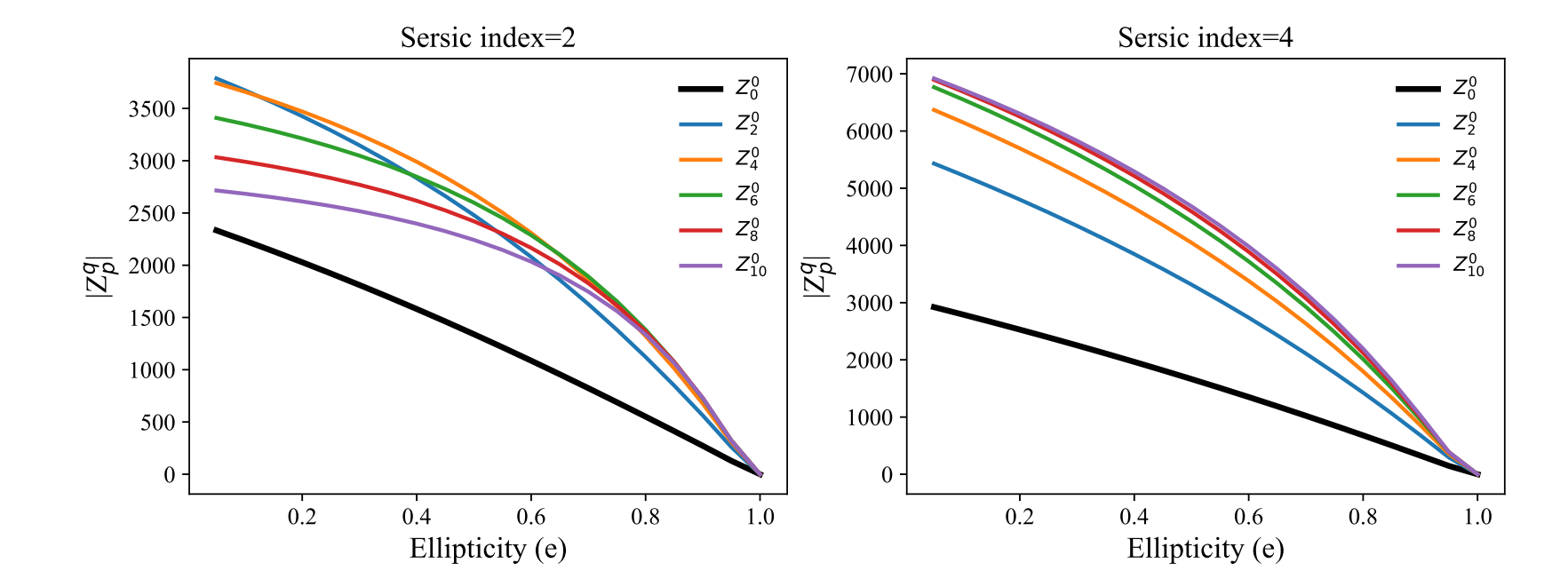}
\caption{ The variation of ZMs $Z_{0}^{0}$ (black line), $Z_{2}^{0}$ (blue line), $Z_{4}^{0}$ (orange line), $ Z_{6}^{0}$ (green line), $Z_{8}^{0}$ (red line), and $Z_{10}^{0}$ (purple line) versus ellipticity parameter of the elliptical galaxies with $n=2$ (left panel) and $n=4$ (right panel). 
 } 
       \label{ellip_Ser2}
\end{figure}

Figure \ref{ellip_Ser2} displays the variation of ZMs $Z_{0}^{0}$ (black line), $Z_{2}^{0}$ (blue line), $Z_{4}^{0}$ (orange line), $ Z_{6}^{0}$ (green line), $Z_{8}^{0}$ (red line), and $Z_{10}^{0}$ (purple line) versus ellipticity parameter of the elliptical galaxies with $n=2$ (left panel) and $n=4$ (right panel). The elliptical galaxies have different shapes, from round (circular) to elongated ellipses, which quantifies with ellipticity parameter (structural) $e$ ranges 0 to 1. As shown in the figure, we observe that the non-vanishing ZMs ($Z_{2}^{0}$) have significant differences at $e=0$ and approach together at $e=1$ for both S\'{e}rsic profiles of $n=2$ and 4. This behavior of the ZPs implies their ability to describe the different elliptical galaxies with different ellipticities.

\begin{figure}
\centering
\includegraphics[width=0.7\textwidth] {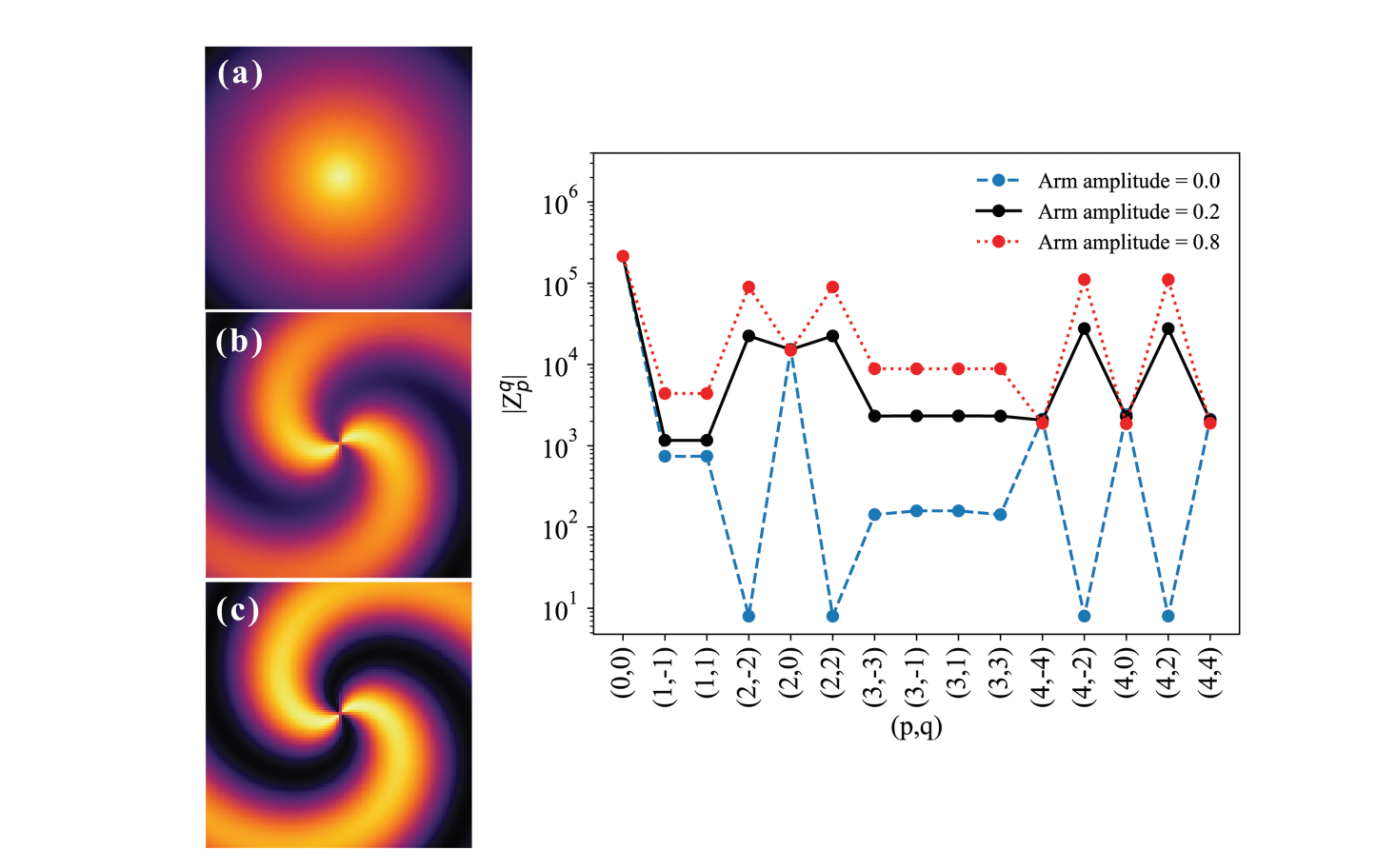}
\caption{ (Left panel) the simulated spiral galaxies with the arm amplitude of 0.0 (a), 0.2 (b), and 0.8 (c), (right panel) the ZMs with $p_{\rm max} = 4$ for each simulated spiral galaxy. The other parameters for three simulated spiral galaxies including the S\'{e}rsic index $n$, effective intensity, effective radius, number of arms, and rotation term are 1, 18.0, 500, 2, and 0.1, respectively.
}
\label{spir}
\end{figure}

 Second, we study the characteristics of ZMs for simulated spiral galaxies. A spiral galaxy is identified by several factors such as bulge, arm (the number, amplitude, and rotation term of arms), disc, and bar \citep{1926ApJ....64..321H}. Figure \ref{spir} shows the simulated spiral galaxies with the arm amplitude of 0.0 (left panel: top), 0.2 (left panel: middle), and 0.8 (left panel: bottom) and the ZMs with $p_{\rm max} = 4$ for each simulated spiral galaxy. For the simulated spiral galaxies, the S\'{e}rsic index, effective intensity, effective radius, number of arms, and rotation term are set to 1, 18.0, 500, 2, and 0.1, respectively. As shown in the figure, the ZMs for $p = 2$ ($Z_{2}^{-2}$, $Z_{2}^{0}$, and $Z_{2}^{2}$) and $p=4$ ($Z_{4}^{-4}$, $Z_{4}^{-2}$, $Z_{4}^{0}$, $Z_{4}^{2}$, and $Z_{4}^{4}$) have different block structures for non-vanishing arm amplitudes. The vanishing arm amplitude indicates the elliptical-like galaxy with $Z_{1}^{q}$, $Z_{3}^{q}$, $Z_{2}^{-2}$, $Z_{2}^{2}$, $Z_{4}^{-4}$, $Z_{4}^{-2}$, $Z_{4}^{2}$, and $Z_{4}^{4}$ are approximately zero. This different behavior of ZMs with different order numbers may help to identify the spiral and elliptical galaxies. 

\begin{figure}
\centering
\includegraphics[width=0.6\textwidth] {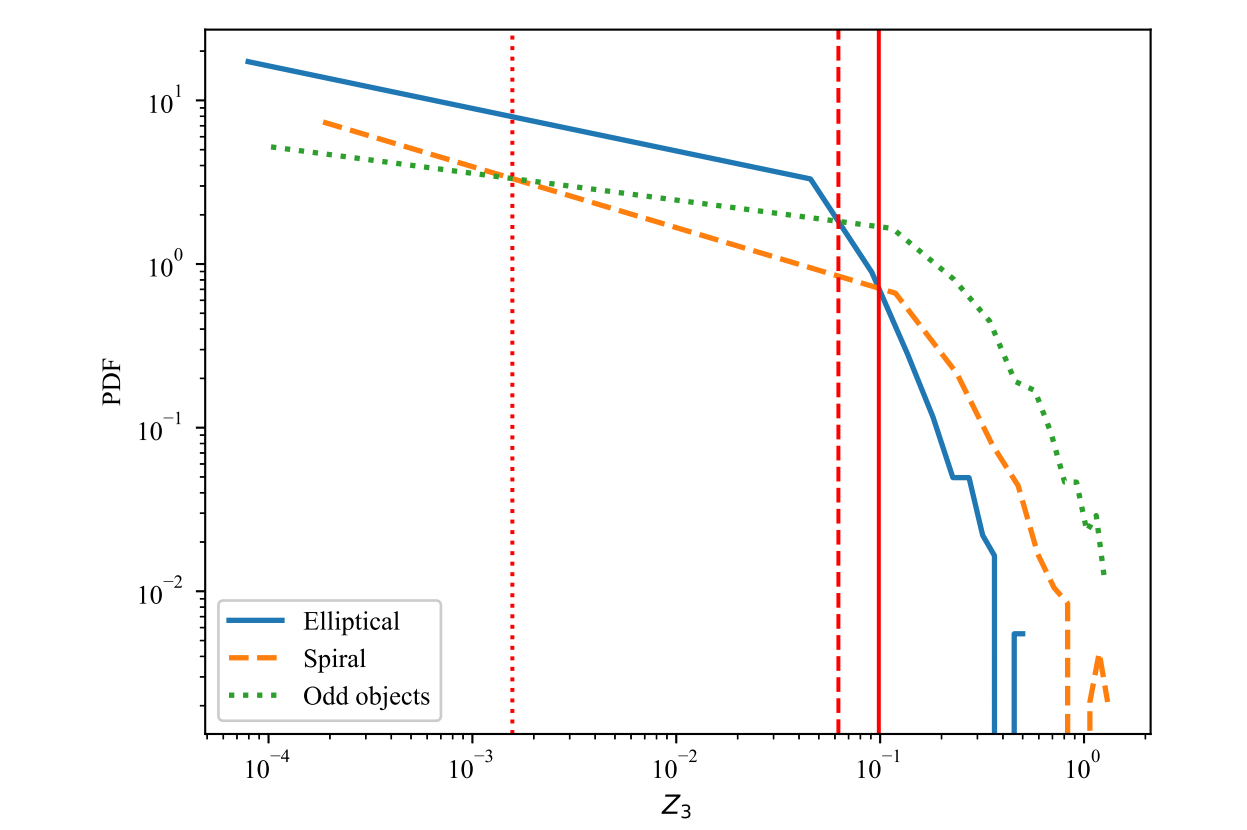}
\caption{ The probability density function (PDF) of the root square ZMs of  $Z_{3}$ for 4077 elliptical (solid blue line), 6139 spiral (dashed orange line), and 1519 odd objects (green dots) collected by GZ2. We illustrate the discriminant boundary between the PDF of elliptical with spiral (vertical solid red line), elliptical with odd objects (vertical dashed red line), and spiral with odd objects (vertical dotted red line).
}
       \label{z3}
\end{figure}

 Figure \ref{z3} represents the probability density function (PDF) of the root square ZMs 
$\sqrt{\sum_{q} Z_{p}^{q}}$ $(q=-3,-1,1,3)$
for 4077 elliptical (solid blue line), 6139 spiral (dashed orange line), and 1519 odd objects (green dots) collected by GZ2. The discriminant boundary (vertical solid red line) between the PDF of spiral and elliptical galaxies indicates the same probability for $Z_{3}$ from both galaxy types. The values of  $Z_{3}$ smaller or larger than the boundary have different probabilities corresponding to the spiral and elliptical galaxies. We observe the similar behavior of $Z_{3}$  PDF discriminating boundaries for elliptical with odd objects (vertical dashed red line) and spiral with odd objects (vertical red dots).
These discriminate boundaries (decision) may help to classify the galaxies based on the $Z_{3}$  values.  For example, for $Z_{3}=0.17$ the PDFs of elliptical, spiral, and odd objects return 0.14, 0.37, and 1.09, respectively. Applying the Bayesian approach \citep[e.g.,][]{Barnes2007SpWea}, we obtain the probability of 0.09 ($\frac{0.14}{0.37+0.14+1.09}$), 0.23 ($\frac{0.37}{0.37+0.14+1.09}$), and
0.68 ($\frac{1.09}{0.37+0.14+1.09}$) correspond to elliptical, spiral, and odd objects, respectively. Our analysis of the PDFs of root square ZMs shows similar behavior for various order numbers. Therefore, ZMs are important descriptors to describe the features of elliptical, spiral, and odd objects.  

\begin{figure}
\centering
\includegraphics[width=0.6\textwidth] {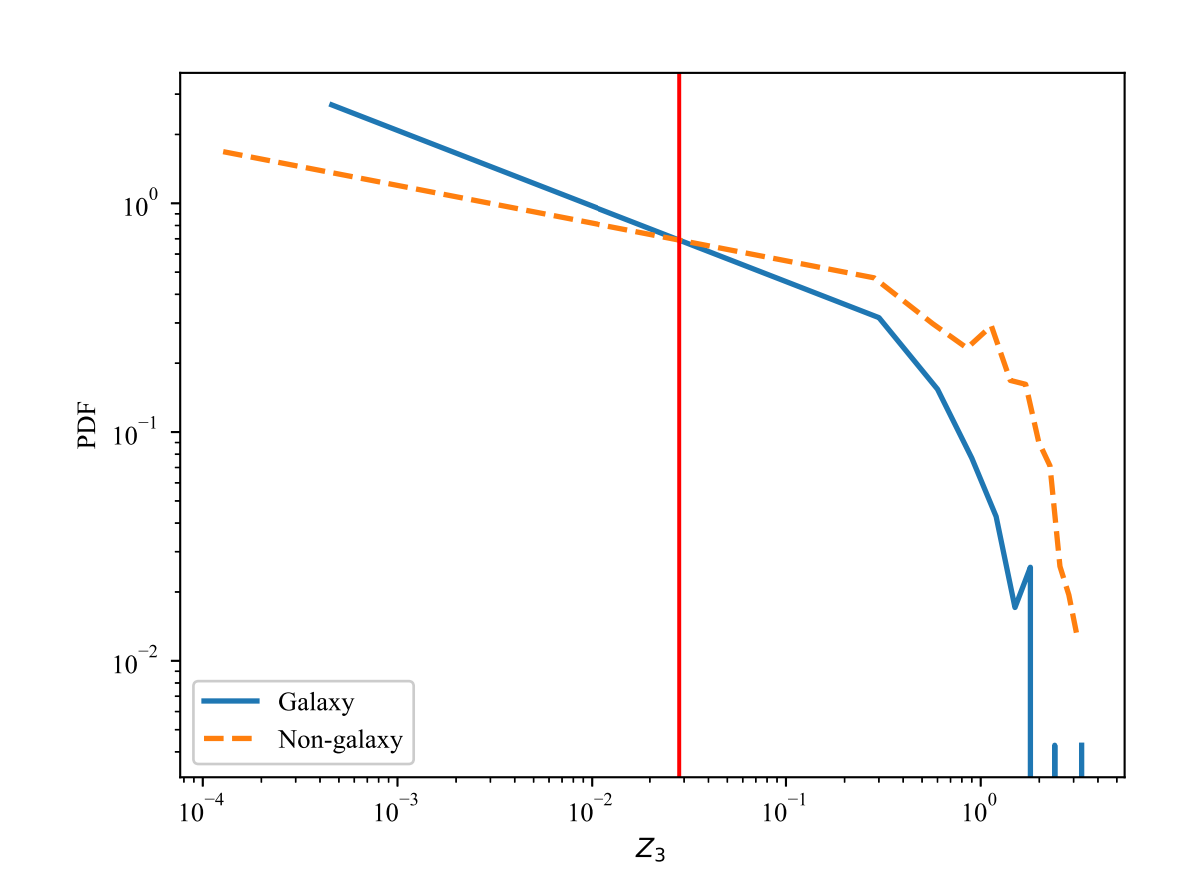}
\caption{The PDF of $Z_{3}$ for 545 non-galaxy images (dashed orange line) and 780 galaxy images (solid blue line). The vertical solid red line indicates the discriminant boundary of $Z_{3}$ for galaxy and non-galaxy samples.
}
       \label{galaxynongalaxy}
\end{figure}

 Figure \ref{galaxynongalaxy} shows the PDF of $Z_{3}$ for 545 non-galaxy images (dashed orange line) and 780 (including 260 samples from each spiral, elliptical, and odd objects) galaxy images (solid blue line). As seen in the figure, the $Z_{3}$ (calculated from ZMs with $p=3$) shows a discriminant boundary (vertical solid red line), implying the capability of ZMs in probabilistic recognizing the galaxies and non-galaxies images in the GZ2 catalog. This probabilistic recognizing capability is also possible for ZMs with different order numbers and discriminant boundaries.  

\begin{figure}
     \centering
     \includegraphics[width=0.6\textwidth] {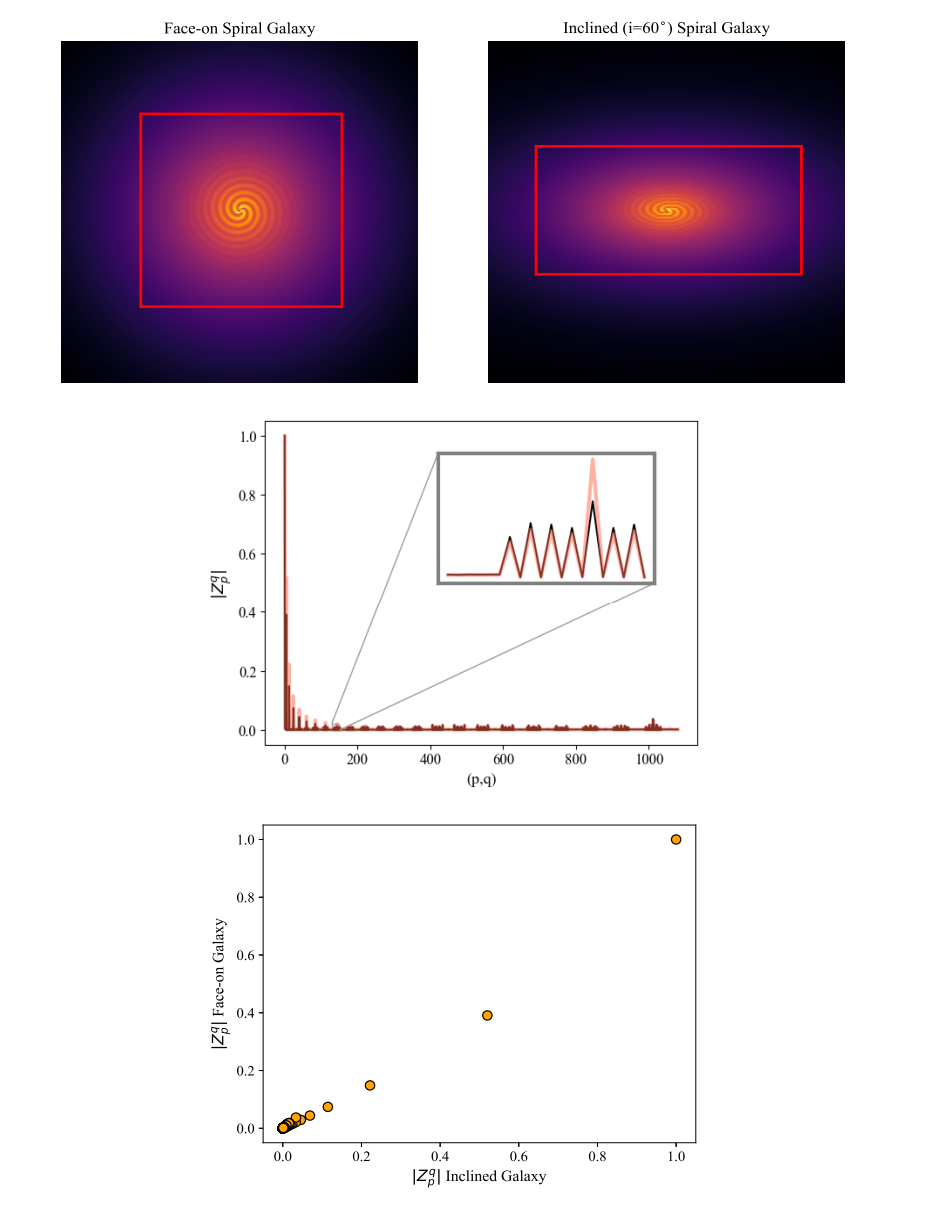}
      \caption{ (top panel) the simulated spiral galaxy profile with S\'{e}rsic index ($n=1$), $R_{e}=100$, and $I_{e}=18$ for inclination angle $i=0^{\circ}$  (left) and $i=60^{\circ}$  (right), (middle panel) the ZMs with $p_{\rm max}=45$ for two perspectives, and (bottom panel) the scatter plot of ZMs for spiral galaxy with $i=0^{\circ}$ (faced-on)  and $i=60^{\circ}$. The inset (middle panel) indicates a large view for a portion of ZMs for two perspectives of the spiral galaxy.}
\label{inclin}  
\end{figure}

 To show the rotation invariant of ZMs, we simulate a spiral galaxy based on the S\'{e}rsic profile with S\'{e}rsic index ($n=1$), $r_{e}=100$, and $I_{e}=18$ observed from two different perspectives with the inclination angle of $i=0 ^{\circ}$  (faced-on) and $i=60 ^{\circ}$. The inclination angle $i$ defines the angle between the plane of the galaxy and the line of sight of the observer, implying the rotation of a galaxy around an axis in its plane.  
Figure \ref{inclin} illustrates a simulated spiral galaxy from two perspectives (top panel), the ZMs (middle panel), and the scatter plot of ZMs (bottom panel). We observe that the block structure of ZMs for two different perspectives with an inclination angle of $i=60^{\circ}$ are similar for different order numbers. Also, the value of ZMs for $i=0^{\circ}$ is strongly correlated with ZMs for $i=60^{\circ}$ (bottom panel). These findings show that the magnitude of ZMs is invariant under rotation (Equation \ref{z_rot}). Therefore, we conclude that the series of ZMs with a maximum order number of $p_{\rm max}$ including a total of  

\begin{equation}
NZMs = \sum_{p=0}^{p_{\text{max}}} (p + 1),
\end{equation}

 ZMs (Figure \ref{recon}) are essential descriptors of galaxy and non-galaxy images that can be classified using machine learning algorithms \citet{SafariIJJA2023}.

\subsection{Galaxy-non-galaxy classifier}\label{primary}

 To identify galaxies and non-galaxies images collected by GZ2 from the SDSS survey, we develop two ZMs-based classifiers (SVM and 1D-CNN), and three original image-based classifiers (2D-CNN, ResNet50, and VGG16) with ViT. The galaxy-non-galaxy classifier uses 545 and 780 non-galaxy and galaxy images, respectively. These 780 galaxy images include 260 samples randomly selected from each 6,139 spiral, 4,077 elliptical, and 1,519 odd objects. The ZMs of images for both galaxies and non-galaxies are computed. Examining the ZMs for maximum order number ($p_{\rm max}$) ranges from 5 to 46, we calculate the performance metric (e.g., accuracy) for both the SVM and classic  1D-CNN classifiers. We obtain the highest performance for $p_{\rm max}=45$ for both SVM and   1D-CNN  models (see the analysis details in Section \ref{secondary}). In the galaxy-non-galaxy classifier, we develop five models identifying galaxy images from the non-galaxy.   We organize five models that are named Model I (SVM with ZMs), II (classic 1D-CNN with ZMs), III (classic 2D-CNN with ViT and original images), IV (ResNet50 with ViT and original images), and V (VGG16 with ViT and original images).  We compare these five models' performance metrics (Table \ref{tab1}) to introduce the highest-performance model. 

\begin{figure}
\centering
\includegraphics[width=0.5\textwidth] {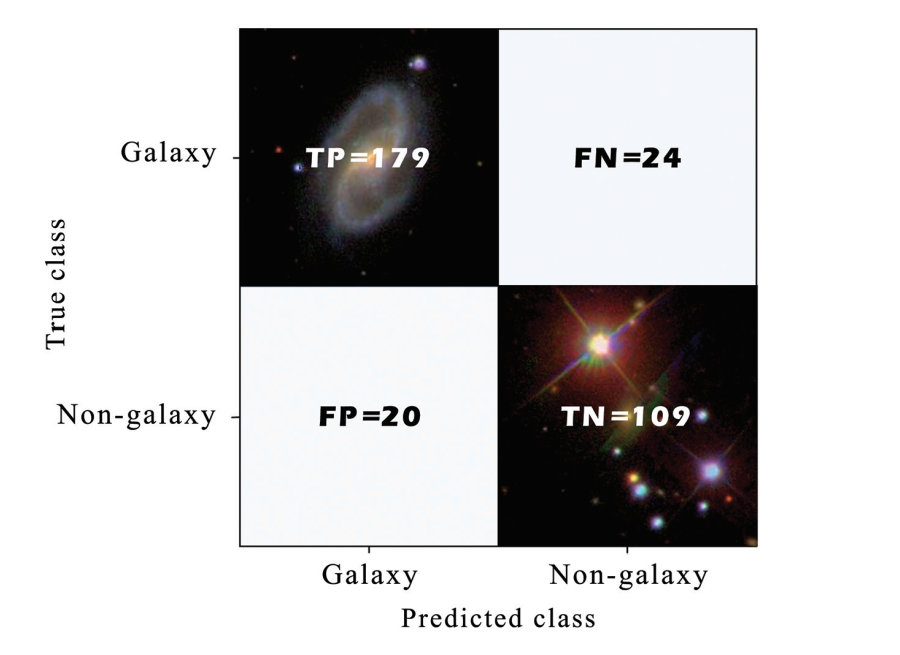}
       \caption{A sample confusion matrix for Model I representing the predicted class vs. true class. True positive (TP) shows the number of galaxies that are correctly classified in galaxy class (positive class). In contrast, false positive (FP) indicates the number of non-galaxy images that are wrongly classified in the galaxy class. True negative (TN) shows the number of non-galaxies that are correctly classified in the non-galaxy class (negative class). In contrast, false negative (FN) indicates the number of galaxy images that are wrongly classified in the non-galaxy class.}
       \label{fig_con}
\end{figure}

Each model uses 75 percent of both classes in the training process, and the remaining 25 percent is applied to the test set. Using the population of the training data set for the positive (galaxy) and negative (non-galaxy) classes, a weighted scalar parameter is set automatically to account for the  imbalanced  class data. Figure \ref{fig_con} represents the confusion matrix for Model I, including true positive (TP), false positive (FP), true negative (TN), and false negative (FN). The TP=179 indicates that Model I correctly classified the number of 179 images in the galaxy (positive) class. At the same time, FN=24 means that 24 images are wrongly galaxy images classified in the non-galaxy (negative) class. The TN=109 shows the Model correctly identified the number of 109 images in the non-galaxy (negative) class. FP=20 implies that 20 images are wrongly non-galaxy images classified in the galaxy (positive) class. The total of TP and FN is equal to the population of the galaxy class. However, the total of TN and FP is equal to the population of the non-galaxy class.

\begin{figure}
\centering
\includegraphics[width=0.7\textwidth] {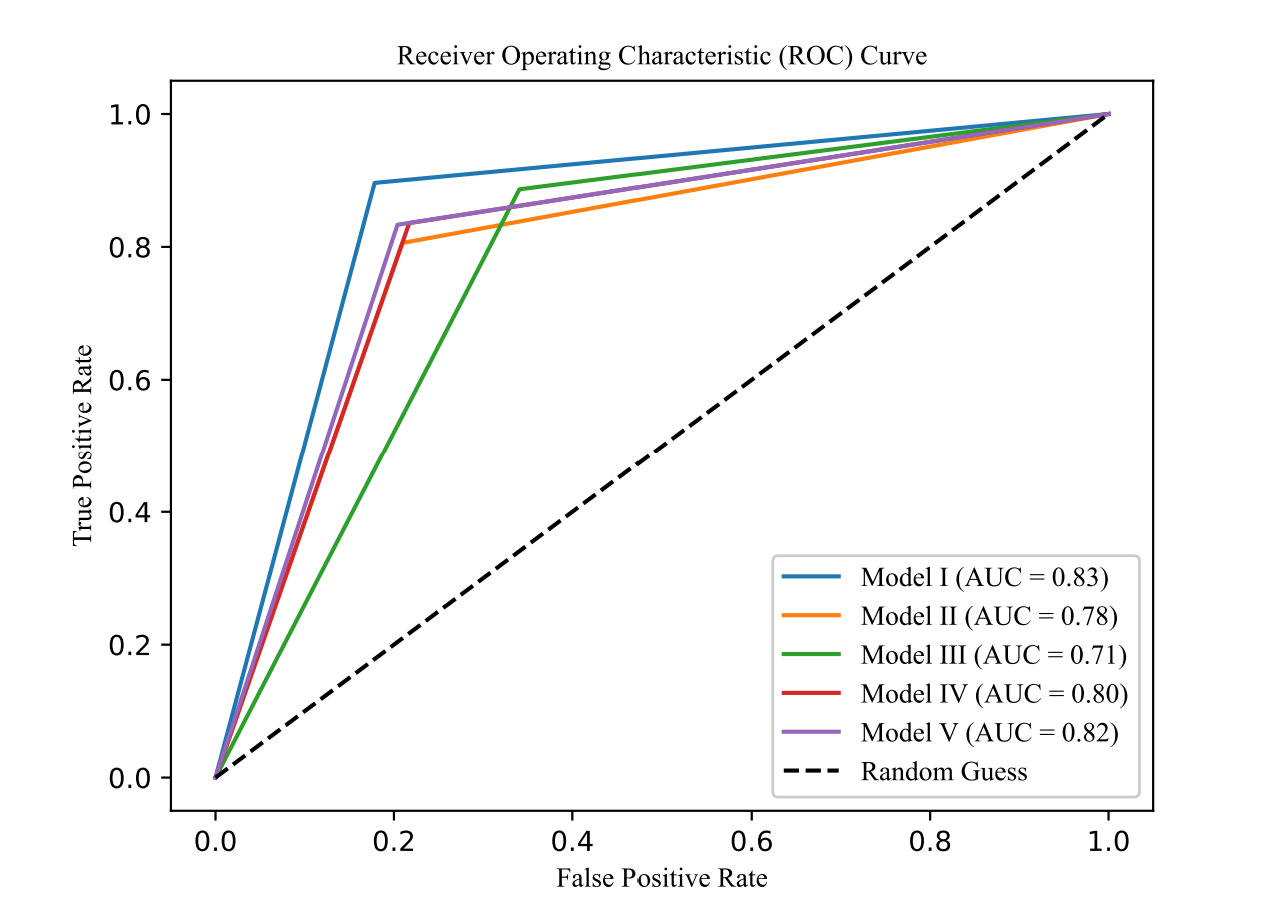}
       \caption{The receiver operating characteristic (ROC) curve for Model I (blue), Model II (orange), Model III (green), Model IV (red), and Model V (purple) of galaxy-non-galaxy classification and random (black dashed line). The area under the curve (AUC) is obtained to be 0.83$\pm$0.02, 0.78$\pm$0.02, 0.71$\pm$0.02, 0.80$\pm$0.02, and 0.82$\pm$0.02 for Model I, Model II, Model III, Model IV, Model V, respectively. }
       \label{roc2}
\end{figure}

A receiver operating characteristic (ROC) curve assesses the performance of binary classifiers (two classes model). The ROC curve shows the true positive rate (TPR)   versus  the false positive rate (FPR) for various classification thresholds that indicate the model's performance to distinguish the positive and negative classes at different levels of certainty. The TPR (FPR) is the proportion of actual positive (negative) that the classifier correctly identifies. Figure   \ref{roc2}  shows the ROC curve for Model I (blue), Model II (orange), Model III (green), Model IV (red), and Model V (purple). The dashed line indicates the ROC curve for the random classifier, comparing the classifier's performance with chance. We   observe  that the ROC curve for the five models is significantly higher than the random's ROC curve, and their performance is much better than chance. Closing the ROC curve to the top left corner of the diagram shows the classifier's performance better than the random classifier. The area under the curve (AUC) indicates the probability that the model will rank a randomly chosen positive sample more significant than a randomly chosen negative case. The AUC is 0.83$\pm$0.02, 0.78$\pm$0.02, 0.71$\pm$0.02, 0.80$\pm$0.02, and 0.82$\pm$0.02 for Model I, II, III, IV, and V, respectively. This implies that all five models classify the galaxies and non-galaxies images much better than the chance. Also, the AUC for Model I and V are more significant than the other three models.

\begin{table}
 \renewcommand{\arraystretch}{1.6}
\vspace{0.5cm}
\caption{ Definition of different skill scores. } \label{tab1} 
\centering
\begin{tabular}{@{}lccccc@{}}
		\hline 
		Score & Formula \\ \hline \hline  
		Recall (positive and negative) & $\rm{recall}^{+} = \dfrac{{\rm{TP}}}{\rm{TP} + \rm{FN}}$ \\ & ${\rm{recall}}^{-} = \dfrac{\rm{TN}}{\rm{TN} + \rm{FP}}$\\[2mm]\hline		
		Precision (positive and negative) & ${\rm{precision}^{+}} = \dfrac{\rm{TP}}{\rm{TP} + \rm{FP}}$\\& ${\rm{precision}^{-}} = \dfrac{\rm{TN}}{\rm{TN} + \rm{FN}}$\\[2mm]\hline
		$f_{1}$ score (positive and negative) & $f_{1}^{+} = \dfrac{2 \times \rm{precision}^{+} \times \rm{recall}^{+} }{\rm{precision}^{+} + \rm{recall}^{+}}$\\ & $f_{1}^{-} = \dfrac{2 \times \rm{precision}^{-} \times \rm{recall}^{-} }{\rm{precision}^{-} + \rm{recall}^{-}}$\\[2mm]\hline
		Accuracy & $\rm{accuracy} = \dfrac{\rm{TP} + \rm{TN}}{\rm{TP} + \rm{FN} + \rm{TN} + \rm{FP}}$ \\[2mm]\hline		
       
       True Skill Statistic (TSS)&$\rm{TSS} = \dfrac{{\rm{TP}}}{\rm{TP} + \rm{FN}} - \dfrac{{\rm{FP}}}{\rm{FP} + \rm{TN}}$\\[2mm]\hline
	\end{tabular}
 \label{tab1}
 \end{table}

Using the elements of the confusion matrix, we calculate the model's performance. The performance of the model classifier can be measured with various metrics such as accuracy, $\rm{precision}^{+}$, $\rm{precision}^{-}$, $\rm{recall}^{+}$, $\rm{recall}^{-}$, ${f_{1}}^{+}$-score, ${f_{1}}^{-}$-score, and TSS. Due to the random selection of train and test datasets, the model metrics fluctuate around their average values. Therefore, we repeated each model's random training and testing more than ten times. Table \ref{tab2} displays the performance metrics of five models. We observe that the accuracy of  Model I (SVM with ZMs) and Model V (VGG16 with ViT and original images)  is slightly greater than that of Model II, III, and IV. The accuracy, precision, and   $f_{1}$-scores  are valuable metrics to measure the performance of class-balanced classifiers; however, the galaxy-non-galaxy classifier is   imbalanced  due to differences in the population of two classes. The TSS is one of the essential scores used to measure the performance of the   imbalanced  classifiers. As shown in the table, the TSS for Model I is the highest  (TSS=0.66$\pm$0.03), while Model III has the lowest value (TSS=0.43$\pm$0.04). Also, the TSS for Model V is in the second rank. Therefore, Models I and V are the best for identifying galaxy-non-galaxy images.

 To obtain the high-performance classification, we examine the SVM and 1D-CNN binary models, including ZMs for 545 non-galaxy images and 11,735 galaxy images (6,139 spiral, 4,077 elliptical, and 1,519 odd objects). The imbalance rate (the ratio of the minority class population to the majority class) is 0.05 for galaxy and non-galaxy classes, showing the high imbalance classification problem. To improve the imbalance rate, the oversampling of the minority class (non-galaxies) is a solution. We add the ZMs of non-galaxy samples for each R, G, and B channel to the training dataset. In this case, the training dataset for the minority class includes the ZMs of the grayscale, R, G, and B channel images, which gives an imbalance rate of about 0.2. Due to differences in the intensity scales of R, G, B, and grayscale images, we normalize the ZMs to the $Z_{0}^{0}$ (total intensity of image) for each image of both training and testing datasets. Please note that the test dataset contains the ZMs of grayscale images for galaxies and non-galaxies samples, which gives an imbalance rate of about 0.05. Indeed, the oversampling does not apply for galaxy samples in the training dataset, so to avoid the high artificial performance metrics for the classifier, we do not use the oversampling for the test dataset. We find 0.90$\pm$0.00, 0.86$\pm$0.01, and 0.93$\pm$0.03 for accuracy, TSS, and AUC, respectively, of SVM with ZMs for galaxy and non-galaxy classifier. Also, we obtain 0.93$\pm$0.01, 0.70$\pm$0.03, and 0.85$\pm$0.01 for accuracy, TSS, and AUC, respectively, of 1D-CNN with ZMs.

\begin{table}
    \centering
    \caption{The performance metrics  for galaxy - non-galaxy classifier are accuracy, $\rm {recall}^{+}$, $\rm{recall}^{-}$, $\rm{precision}^{+}$, $\rm{precision}^{-}$, ${f_{1}}^{+}$, ${f_{1}}^{-}$, and True Skill Statistic (TSS). The mean values and standard deviations of metrics are obtained for ten iterations and determined from five classification models.} 
    \label{tab2}
    \begin{tabular}{@{}lccccc@{}}
        \toprule
        Score & Model I & Model II & Model III & Model IV & Model V \\
        \hline
        \midrule
        $\rm{Recall}^{+}$        & 0.90$\pm$0.01 & 0.83$\pm$0.05 & 0.93$\pm$0.08 & 0.85$\pm$0.05 & 0.86$\pm$0.05 \\
        $\rm{Recall}^{-}$        & 0.76$\pm$0.03 & 0.76$\pm$0.06 & 0.50$\pm$0.09 & 0.76$\pm$0.07 & 0.78$\pm$0.06 \\
        
        $\rm{Precision}^{+}$     & 0.85$\pm$0.02 & 0.83$\pm$0.03 & 0.72$\pm$0.02 & 0.84$\pm$0.03 & 0.85$\pm$0.03 \\ 
        $\rm{Precision}^{-}$     & 0.84$\pm$0.02 & 0.76$\pm$0.05 & 0.86$\pm$0.09 & 0.79$\pm$0.05 & 0.80$\pm$0.06 \\
        
        ${f_{1}}^{+}$   & 0.87$\pm$0.01 & 0.83$\pm$0.02 & 0.81$\pm$0.03 & 0.84$\pm$0.02 & 0.86$\pm$0.02 \\
        ${f_{1}}^{-}$   & 0.80$\pm$0.02 & 0.76$\pm$0.02 & 0.62$\pm$0.05& 0.77$\pm$0.03& 0.79$\pm$0.02 \\
        
        $\rm{Accuracy}$          & 0.84$\pm$0.01 & 0.81$\pm$0.01 & 0.75$\pm$0.02 & 0.81$\pm$0.02 & 0.83$\pm$0.02 \\
        TSS                      & 0.66$\pm$0.03 & 0.60$\pm$0.03 & 0.43$\pm$0.04 & 0.61$\pm$0.04 & 0.64$\pm$0.04 \\
        \bottomrule
    \end{tabular}
    \label{tab2}
\end{table}

\subsection{Galaxy classifier}\label{secondary}

\begin{figure}
\centering
\includegraphics[width=0.8\textwidth] {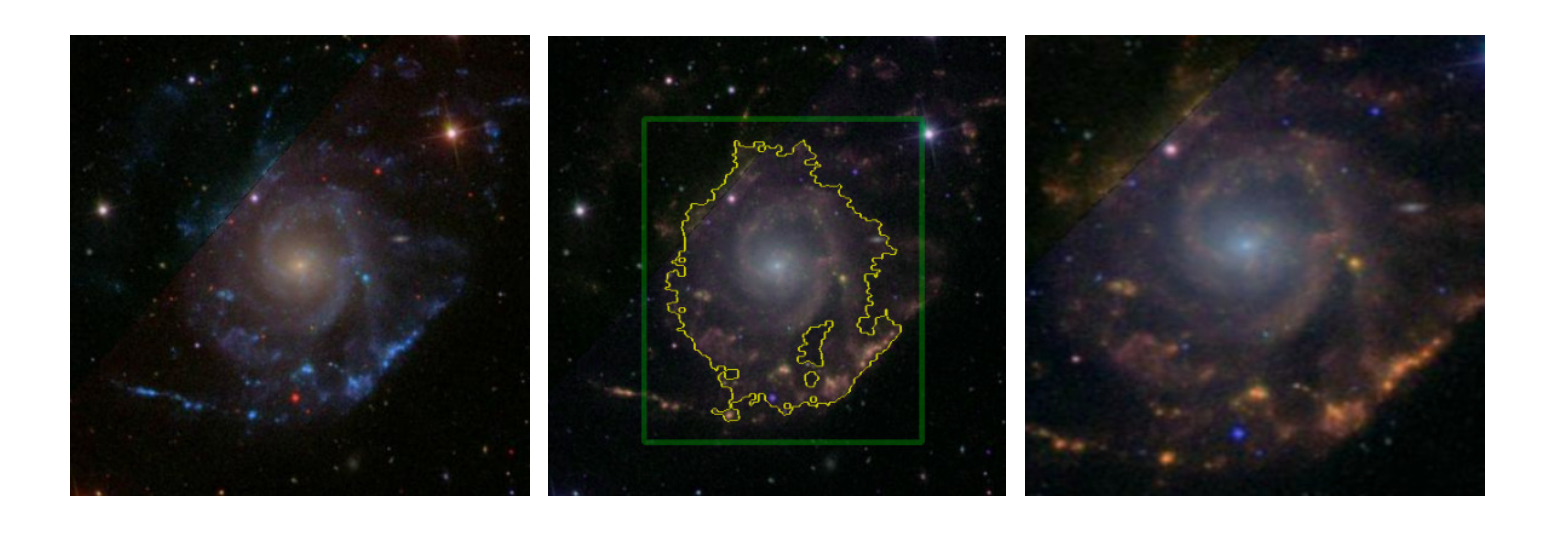}
       \caption{The original image recorded by SDSS (left panel), segmented galaxy represented within a green rectangle (middle panel), galaxy sub-image (right panel).}
       \label{segm}
\end{figure}

The galaxy classifier is designed to classify the galaxy images into three main groups: spiral, elliptical, and odd objects. We collected the population of each class from  gz2$\_$hart16  by applying tasks for classes. For adequate analysis, first, we   apply  a segmentation algorithm (watershed) to each galaxy image. Determining the border of the central event (galaxy or primary feature), we selected a squared   sub-image  slightly more significant than the galaxy border.  In the segmentation process, we transform the central sub-image to the central objects (galaxy) which is called image normalization \citep{Raboonik2017ApJ}. 
The final   sub-image  is resized to   200$\times$200  pixels. Figure \ref{segm} shows an original GZ2 image for a spiral galaxy (left panel), segmented image (middle panel), and galaxy sub-image (right panel). Now, we  compute  the ZMs for each   sub-image  of all galaxy samples. The samples are recognized in three galaxy classes using five classification models (Model I, II, III, IV, and V). Models I and II use the ZMs of cropped images, while Models III, IV, and V perform on the original cropped image for galaxies.

\begin{figure}
\centering
\includegraphics[width=0.5\textwidth] {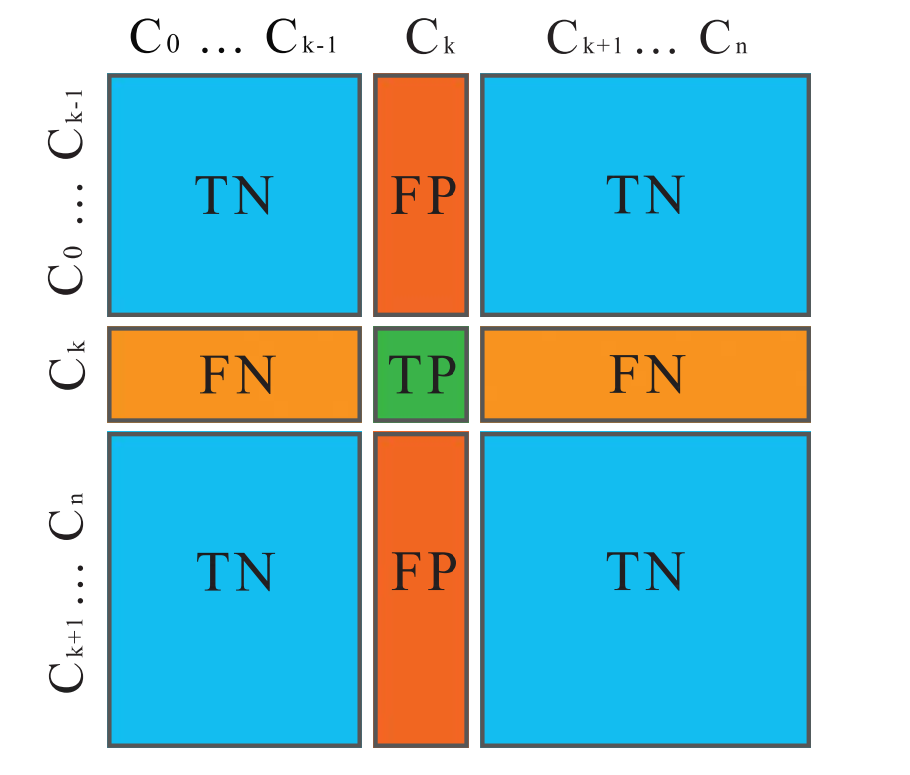}
       \caption{The schematic table of the confusion matrix represents the actual label vs predicted labels for a three-class model.}
       \label{cm3}
\end{figure}

To evaluate the performance of these five classification models, we need to make the confusion matrix the foundation for most evaluation metrics (Table \ref{tab1}). Similarly to the two-class model, the confusion matrix of three or multi-class classification contains rows including authentic (actual) classes and columns including predicted classes. Its entries represent how many instances are classified into each class compared to how the model predicted them. Figure \ref{cm3} shows a 3$\times$3 schematic table with rows including the actual classes and columns including the predicted classes. 

\begin{figure}
\centering
\includegraphics[width=0.9\textwidth] {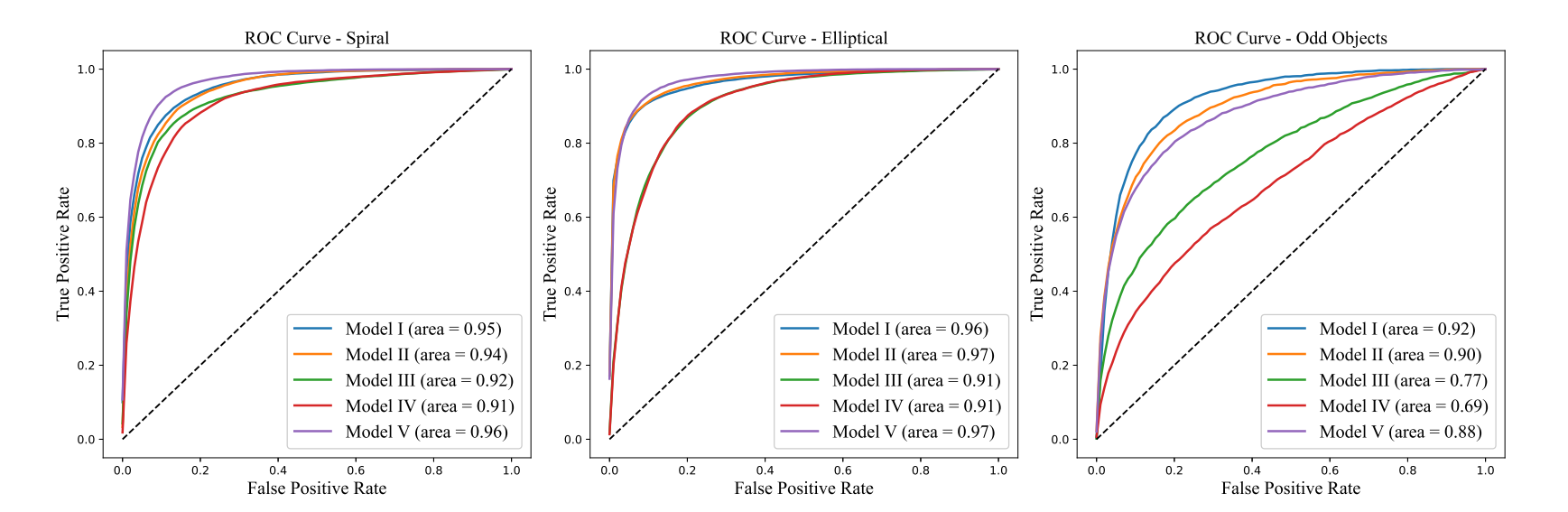}
       \caption{The receiver operating characteristic (ROC)  curve for spiral (left panel), elliptical (middle panel), and odd objects (right panel) classes of Model I (blue), Model II (orange), Model III (green), Model IV (red), and Model V (purple), and random (black dashed line). The area under the curve (AUC) is obtained for each class and every model.}
       \label{roc3}
\end{figure}

Figure   \ref{roc3}   displays the ROC curve for spiral (left), elliptical (middle), and odd objects (right) classes for Model I (blue), Model II (orange), Model III (green), Model IV (red), and Model V (purple), and random (black dashed line). The value of AUC for spiral and elliptical classes of five classification models is more significant than 0.9, indicating the ability of these models to identify the spiral and elliptical galaxies. While the AUC for odd   objects  is less than 0.9 for Models III, IV, and V, designed based on original images, the AUC is more than 0.9 for Models I and II that used the ZMs. 

Table \ref{tab3} tabulates the performance metrics for five galaxy classification models. The overall accuracy metric is more significant than 0.9 for Models I, II, and V. The recall score represents the true positive rate of a classification model obtained about 0.84 for odd objects in Model I (uses SVM   with  ZMs), which is greater than the other four models. However, the precision score is about 0.81 for odd objects in Model V (uses   VGG16 with ViT and original image). The population of the odd objects from GZ2 in our data set is at least one-third of spiral and elliptical galaxies. Hence, the galaxy classification problem is   imbalanced  data sets where spiral and elliptical classes have significantly more samples than the class of the  odd objects.

One solution to measure an  imbalanced  class model's performance is calculating the weighted metrics that apply to the population of classes and metric values in a multi-class model. The weighted recall and precision are obtained slightly more significantly than 0.9 for Models I, II, and V, implying the goodness of the classification of galaxies via these three models. However, in the context of measuring classification algorithms' performance, the TSS is a suitable metric for studying the performance of  imbalanced  class models. The TSS is greater than 0.86 for Models I, II, and V. Although, the performance metrics of  SVM with ZMs and 1D-CNN with ZMs  classification models are slightly similar to the  VGG16 with ViT and original image. 

\begin{table}
    \centering
    \caption{ The performance metrics (recall, precision, $f_{1}$-score) for spiral, elliptical, and odd objects galaxy classes. Accuracy is the most fundamental metric that indicates the overall proportion of correctly classified instances for each classification model. The overall metrics (recall, precision, $f_{1}$-score, and TSS) are weighted for each classification model. The mean values and standard deviations of metrics are obtained for ten iterations and determined from five classification models. } 
    \begin{tabular}{ccccccc}
        \toprule
        Score & Class/Overall & Model I & Model II & Model III & Model IV & Model V \\
        \hline
        \midrule   
        
       \multirow{4}{*}{\rm\bf{{Recall}}}  
       
       &\rm{Spiral}& 0.86$\pm$0.01 & 0.89$\pm$0.03 & 0.90$\pm$0.05 &0.88$\pm$0.05 & 0.95$\pm$0.01\\   
       
       & \rm{ Elliptical}& 0.99$\pm$0.003 & 0.99$\pm$0.006 & 0.92 $\pm$ 0.08 &0.64$\pm$0.20& 0.98$\pm$0.01\\ 
       
       &\rm{Odd {\bf objects}}& 0.84$\pm$0.02 & 0.71$\pm$0.08 & 0.50 $\pm$ 0.08 &0.26$\pm$0.14& 0.70$\pm$0.05\\    
       
       &\rm{Weighted}& 0.91$\pm$0.004 & 0.91$\pm$0.01 & 0.88 $\pm$ 0.03 &0.79$\pm$0.04& 0.94$\pm$0.003\\
       \midrule 

       \multirow{4}{*}{\rm\bf{{Precision}}}  
       
       &\rm{Spiral}& 0.96$\pm$0.006 & 0.93$\pm$0.02 & 0.88$\pm$0.03 &0.78$\pm$0.07& 0.93$\pm$0.01\\   
       
       & \rm{ Elliptical}& 0.99$\pm$0.002 & 0.98$\pm$0.01 & 0.90$\pm$0.05 &0.84$\pm$0.10& 0.95$\pm$0.01\\ 
       
       &\rm{Odd objects}& 0.60$\pm$0.02 & 0.64$\pm$0.06 & 0.64$\pm$0.10 &0.37$\pm$0.24& 0.81$\pm$0.03\\   
       
       &\rm{Weighted}& 0.93$\pm$0.004 & 0.92$\pm$0.005 & 0.86$\pm$0.02 &0.78$\pm$0.04& 0.93$\pm$0.003\\
       \midrule 

       \multirow{4}{*}{\rm\bf{$f_{1}$}} 
       
       &\rm{Spiral}& 0.91$\pm$0.005 & 0.91$\pm$0.01 & 0.88$\pm$0.02 &0.82$\pm$0.03& 0.94$\pm$0.003\\ 
       
       & \rm{ Elliptical}& 0.99$\pm$0.002 & 0.98$\pm$0.004 & 0.90$\pm$0.03 &0.70$\pm$0.17& 0.97$\pm$0.004\\ 
       
       &\rm{Odd objects}& 0.70$\pm$0.01 & 0.67$\pm$0.03 & 0.54$\pm$0.05 &0.22$\pm$0.10& 0.74$\pm$0.02\\  
       
       &\rm{Weighted}& 0.91$\pm$0.004 & 0.91$\pm$0.01 & 0.86$\pm$0.02 &0.77$\pm$0.05& 0.93$\pm$0.003\\
       \midrule
       
       \rm\bf{Accuracy}        
       
       && 0.90$\pm$0.005 & 0.90$\pm$0.01 & 0.85$\pm$0.03 &0.72$\pm$0.10& 0.93$\pm$0.003\\
       \midrule  
       
       \rm\bf{TSS}        
       
       &\rm{Weighted}& 0.88$\pm$0.003 & 0.86$\pm$0.01 & 0.77$\pm$0.04 &0.61$\pm$0.07& 0.89$\pm$0.005\\ 
  
    \bottomrule
    \end{tabular}
    \label{tab3}
\end{table}

\begin{figure}
\centering
\includegraphics[width=0.7\textwidth] {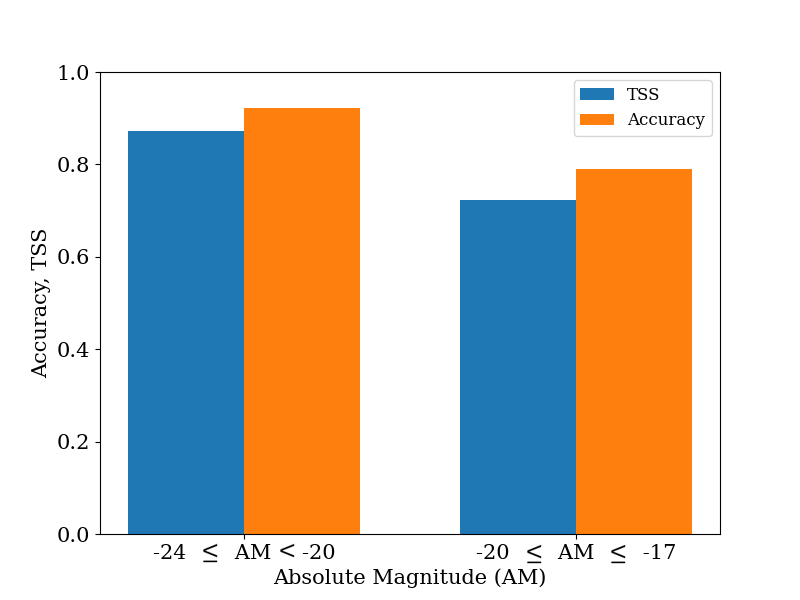}
\caption{ Variation of Accuracy (orange) and TSS (blue) versus absolute magnitudes of galaxies (AM).
}
\label{absmag}
\end{figure}

Figure \ref{absmag} displays the variation of global performance metrics of accuracy and TSS versus the absolute magnitude (Petrosian absolute magnitude in r-band) of galaxy samples obtained by the SVM with the ZMs classifier. We find that the SVM with ZMs well classified the galaxies into spiral, elliptical, and odd objects with accuracy more prominent than 0.79 and TSS larger than 0.72 for the absolute magnitude greater than -20. However, for absolute magnitude less than -20, the accuracy and TSS are about 0.92 and 0.87, respectively. We obtain similar results for 1D-CNN with ZMs and absolute magnitudes. 

\section{Discussion}\label{disc}
The ZMs with $p_{\rm max}=45$ give 1081 independent and unique features describing a galaxy image (Figure \ref{recon}) that can reconstruct the original image with a minimum reconstruction error. Also, the ZMs show many similarities in their structures for the digital photos of a class, which helps with high-performance classification based on machine learning. For example, we find that the PDF of ZMs with various order numbers for non-galaxies, spirals, ellipticals, and odd objects indicate the discriminant boundaries (Figures \ref{z3} and \ref{galaxynongalaxy}), showing the differences of block structures for different galaxy classes. However, these block structures show similar trends (local patterns) with slight differences in the ZMs magnitudes for the galaxies of the same types. For instance, the ZMs' block structure for $p=3$ includes  $Z_{3}^{-3}$, $Z_{3}^{-1}$,$Z_{3}^{1}$, and $Z_{3}^{3}$ is different for spiral, elliptical, and odd objects due to the differences in their inherent shape characteristics. Indeed, the ZMs with $p=3$ calculate integration over the phase terms of $\cos(|q|\theta)$ and $\sin(|q|\theta)$ ($q=$-3, -1, 1, and 3) for galaxies returning different values due to, e.g. axially symmetry for ellipticals and arms shapes for spirals. The magnitude of ZMs for a galaxy from different perspective angles is invariant (Figure \ref{inclin}), indicating the rotation invariances of ZMs. Indeed, ZMs are important descriptors for interpreting the surveys' observed galaxies with various inclination angles. We show that the ZMs components are sensitive to the structural and morphological properties that can help to interpret the galaxy images. Rotation invariant properties are essential concepts for galaxy classification and morphological prediction in the machine learning framework \citep[e.g.,][]{Dieleman2015MNRAS}. The efficiency and performance of galaxy classification depend on the feature selection process from galaxy images. Obtaining the one-dimensional features from galaxy images based on elliptical isophote analysis is a solution for feature selection for machine learning \citep{ Tarsitano2022MNRAS.511.3330T}.

Feature selection is an essential step in machine learning due to improving model performance, reducing overfitting and noises, enhancing interpretability, lowering computational cost, and mitigating the curse of dimensionality \citep{kohavi1997Elsevier, Guyon2003JMLR}. Machine learning models such as 2D-CNN, ResNet50, and VGG16 with ViT are among the image classifiers that extract features based on activation mapping that depends on the number of channels and size of the filters in  convolutional  layers (Figure \ref{cnnscheme}). The length of the final extracted features in such machine learning algorithms depends also on stride, padding, and pooling operations. For example, of a typical galaxy image with the size of 200$\times$200 pixels, the VGG16 with 13, 5, and 3 of convolutional, max-pooling, and fully-connected layers, respectively, computing 138 million trainable parameters retains 25,088 final features in the last max-pooling layer to describe the image. However, the independence and uniqueness of these features are challenging, but the VGG16 learning model follows the similarities in the features of each class to classify images of different classes. The independence and uniqueness of ZM features for galaxy data sets may reduce the overfitting and redundant information, hence achieving lower computational cost and higher performance. 

We show that the performance metrics (Table \ref{tab2}) of the SVM with ZMs for a binary class of 545 non-galaxies and 780 galaxies are slightly more significant than the other four models (1D-CNN with ZMs, 2D-CNN, ResNet50, and VGG16 with ViT). The SVM with ZMs uses 1081 features for each image, but the three imaged-based classification models work with many features. This implies the less computational cost for SVM with ZMS. Therefore, we increase the number of galaxy classes to 11,735 and apply the oversampling approach for non-galaxy classes (minority class). Increasing the number of galaxy samples, the SVM or 1D-CNN can determine enough patterns and structures in the ZMs for galaxies, leading to precisely identifying them with non-galaxy features that include stars, artifacts, cosmic rays, image noise, and diffraction spikes \citep{Willett_2013}. The oversampling of the minority class adds the ZMs of the R, G, and B channels, which returns an imbalance rate of about 0.2.
The oversampling approach is a solution to increase the population of the minority class in the case of the unavailability of more new samples for non-galaxy in the  GZ2 catalog. Indeed, the GZ2 is mainly a project for galaxies, so the non-galaxy samples are less frequent in this catalog. In this case, SVM can separate ZMs of galaxies and non-galaxies by applying the decision boundary (hyperplane). At the same time, the 1D-CNN transforms input ZMs through layers of learned filters and makes decisions based on high-dimensional, hierarchical features. The TSS and AUC in the ROC curve for SVM with ZMs of the final galaxy-non-galaxy increase to 0.86 and 0.93, respectively. Expectedly, these results are obtained with a lower computational cost for SVM and 1D-CNN algorithms that used the one-dimensional ZMs in feature space. The GitHub codes include the classification process for SVM and 1D-CNN, which is entirely runnable on fewer CPU processors (less computational cost). However, each 2D-CNN, ResNet50, and VGG16 needs remarkable GPUs to process many original-size 424$\times$424$\times$3 galaxy and non-galaxy images. Previously, the algorithms for classifying galaxies and star sub-images were investigated for other databases. \citet{Baqui2021A&A} developed the machine learning-based star (point-like features) and galaxy classification algorithm using photometric and morphological features for the miniJPAS catalog of more than 64,000 detection band samples. They obtained the AUC=0.95 and 0.98 for random forest and highly randomized tree classifiers, respectively. \citet{Stoppa2023A&A} applied star-galaxy classification using a CNN algorithm with unique information for MeerLICHT telescope images and using the Dark Energy Camera Legacy Survey (DECaLS). They obtained the classifier's performance (ROC-AUC, PR-AUC, and Brier score) for various signal-to-noise ratios above the detection limit.

Applying the five models to classify galaxy samples into spiral, elliptical, and odd objects indicates that the performances of ZMs-based models (SVM and 1D-CNN) are comparable with the VGG16 with ViT and original images. The global metrics, including accuracy and TSS, are greater than 0.9 and 0.86, respectively. The one-dimensional ZMs-based algorithms use 1081 features for each sample in the classification process, while the pre-trained VGG16 deep learning model uses more than 25,000 features, reducing from 138 million trainable parameters. We find that both SVM and 1D-CNN  (considering the computing of the ZMs and train-test process)  need lower Computational Process Units (CPUs) compared with VGG16 with ViT, which requires considerable Graphical Process Units (GPUs). 

\citet{freed2013application} applied an SVM model to classify spiral, elliptical, and irregular galaxies. They used an equal size of 1000 samples in each class for training and test sets. They reported the accuracy for the galaxy classifier. Using the ResNet classifier, \citet{Zhu2019Ap&SS} obtained an accuracy of about 95 percent for the morphology classification of galaxies. \citet{Gupta2022A&C} performed the NODE method for galaxy classification, obtaining an accuracy of around 91-95 percent. Their method advantage was using fewer parameters, hence less computational memory. \citet{Walmsley2020MNRAS} applied a Bayesian CNN algorithm for morphological classification galaxies of the GZ catalog and obtained considerable performance compared with other approaches. \citet{Li2022MNRAS} investigated a hybrid model (RegNetX-CBAM3) for classifying GZ2 and   EFIGI  data sets into seven classes. They reported 92.02, 92.14, 92.13, 92.10, and 98.27 for accuracy, purity, $f_{1}$-score, and AUC, respectively, for the best training data set validated for the model classifier. 
They performed the galaxy classification using several algorithms. They obtain the accuracy of best training data set of galaxy classification about 84.34, 84.63, 85.57, 87.45, 88.74, 88.80, 89.68, 90.06, 90.06, and 91.44 for AlexNet \citep{krizhevsky2012}, MobileNetV2 \citep{SandlerMark}, DenseNet \citep{HuangGao}, GoogLeNet \citep{SzegedyChristian}, ShuffleNetV2 \citep{ma2006ubiquitous}, ResNet \citep{HeKaiming}, EfficientNetV2 \citep{tan2021efficientnetv2}, EfficientNet, \citep{tan2019efficientnet}, VGG16 
\citep{Simonyan2014arXiv}, and
RegNetX \citep{Radosavovic_2020_CVPR}, respectively compared with the RegNetX-CBAM3 classification model. \citet{Cao2024A&A} achieved significant accuracy in the morphological classification of galaxy images by applying a convolutional vision transformer algorithm on the GZ dataset.

\section{Conclusion}\label{con}

We used five classification models to classify galaxy images. First, we collected a sample of 545 non-galaxy images and 11,735 galaxy images from the GZ2 catalog. To do this, we applied the task answer of GZ2 via the fraction thresholds to recognize the samples mentioned above. Second, we computed the ZMs for all images. The ZMs   represent unique  and independent features   that describe  image structures and shapes, which   are  essential properties for galaxies due to their circular-like shapes. We observed that using ZMs, the reconstructed image of the galaxy returns the original image's structures, implying the efficiency of the finite number of ZMs to describe the galaxy images (Figure \ref{recon}). We showed that the ZMs have different block structures (for each order number) for spiral and elliptical galaxies due to arm structures for spirals and axial symmetry   properties for ellipticals. The ZMs with even order numbers ($p = 2l$) and $q=0$ were indicators for elliptical galaxies, while the ZMs with odd numbers showed different block structures of spirals compared with ellipticals. We obtained the discriminant boundaries for ZMs with different order numbers of spiral, elliptical, and odd objects, indicating the probabilistic classification of galaxies in the Bayesian framework (Figures \ref{z3} and \ref{galaxynongalaxy}). We showed that the magnitude of ZMs is invariant under rotation (from two perspectives of inclination angles) due to the exponential term in the ZPs (Figure \ref{inclin}) that satisfied  Equation \ref{z_rot}. These unique characteristics (block structures and decision boundaries) of ZMs for different types of galaxies and invariances (under rotation, translation, and scaling) make   ZMs essential features to feed statistical machine learning algorithms  to classify GZ2 images. We applied the segmentation algorithm to extract a central galaxy from GZ2 images, which held the central sub-image in the center of brightness.   This transformation normalizes galaxy images, ensuring translation and scale invariance of the ZMs.

 In statistical machine learning, the number of employed features can affect the classification accuracy and the computational cost.  A total of 1081 ZMs for maximum order number $p_{\rm max}=45$ for each image was used to design the classification  algorithm  comparing with the total of more significant than 3, 25.6, and 138 million parameters applied in 2D-CNN (3 convolutional layers), ResNet50 (50 convolutional layers), and VGG16 (13 convolutional layers) for each original image.  In the case of 545 non-galaxy images and 780 galaxy images, the various performance metrics (such as accuracy, precision, recall, $f_{1}$-score, and TSS) were computed by applying the elements of the confusion matrix for five models. We observed that using ZMs in the SVM machine (Model I) gives a high-performance classification for our data set, recognizing galaxies from non-galaxy images. The TSS of Models I and V was about 0.64 (Table \ref{tab2}), and their AUC of the ROC curve was more significant than 0.82 (Figure \ref{roc2}). The AUC of the ROC curve indicates the ability of the classifier to compare with the random (chance) model.  We increased the number of galaxy samples to 11,735, and after applying the oversampling process for non-galaxy samples, the TSS for the SVM with Zernike Moments (ZMs) rose to 0.86.

To identify the galaxy images into spiral, elliptical, and odd objects, we first applied a segmentation algorithm to select the central object from galaxy images. The ZMs of segmented galaxy images were computed. Then, we applied the five classification models to identify galaxy images to spiral, elliptical, and odd objects. We calculated the weighted performance metrics for three-class models. The TSS was obtained 0.86, 0.88, and 0.89 for SVM with ZMs, 1D-CNN with ZMs, and VGG16 with ViT, respectively (Table \ref{tab3}). We obtained that the SVM and 1D-CNN with ZMs classified faint and bright galaxies with different absolute magnitudes. However, the classification performance for bright galaxies was slightly more significant than faint galaxies collected from the GZ2 catalog (Figure \ref{absmag}).  

 Our analysis showed that the SVM and 1D-CNN with ZMs for classifying galaxies are efficient and use less CPU and GPU in training and testing. Increasing the data quality and quantity of modern observations, the feature selection based on ZMs can describe galaxy and non-galaxy objects. Due to the characteristics of ZMs such as invariancy, uniqueness, and completeness, are the utility of ZMs to describe images which is essential to avoid overfitting in the machine learning models. The ZMs-based classification algorithms for galaxies and non-galaxies objects may help to improve the imbalance rate classification of galaxies and non-galaxies. Therefore, the ZMs-based classifying algorithms are a solution for recognizing billions of galaxies and objects from near-future stage-IV observations.  

\clearpage
\section*{Acknowledgements}
We gratefully thank the
anonymous referee for very helpful comments and suggestions
to improve the manuscript.

\bibliographystyle{apj}
\bibliography{MS.bib} 

\end{document}